\documentclass{jpp}
\usepackage{graphicx}

\usepackage[utf8]{inputenc}
\usepackage[T1]{fontenc}
\usepackage{amsmath}

\usepackage{color,soul} 
\usepackage{comment}

\usepackage{xcolor} 
\usepackage{ulem} 

\title{Kinetic Entropy-Based Measures of Distribution Function Non-Maxwellianity: Theory and Simulations}

\author{Haoming Liang\aff{1,2}
  \corresp{\email{haoming.liang@uah.edu}}, M.~Hasan~Barbhuiya\aff{2}, P.~A.~Cassak\aff{2,3}, O. Pezzi\aff{4,5}, S.~Servidio\aff{6}, F.~Valentini\aff{6}, G.~P.~Zank\aff{1,7}}

\affiliation{\aff{1}Center for Space Plasma and Aeronomic Research, University of Alabama in Huntsville, Huntsville, AL 35899, USA
\aff{2}Department of Physics and Astronomy, West Virginia University, WV, 26506, USA
\aff{3}Center for KINETIC Plasma Physics, West Virginia University, WV, 26506, USA
\aff{4} Gran Sasso Science Institute, Viale F. Crispi 7, I-67100 L’Aquila, Italy
\aff{5} INFN/Laboratori Nazionali del Gran Sasso, I-67100 Assergi (AQ), Italy
\aff{6}Dipartimento di Fisica, Universit\`a della Calabria, I-87036 Rende (CS), Italy
\aff{7}Department of Space Science and Center for Space Plasma and Aeronomic Research, University of Alabama in Huntsville, Huntsville, AL 35899, USA}

\begin{document}

\maketitle

\begin{abstract}
We investigate kinetic entropy-based measures of the non-Maxwellianity of distribution functions in plasmas, i.e., entropy-based measures of the departure of a local distribution function from an associated Maxwellian distribution function with the same density, bulk flow, and temperature as the local distribution. First, we consider a form previously employed by Kaufmann and Paterson [{\it J.~Geophys.~Res.,} {\bf 114}, A00D04 (2009)], assessing its properties and deriving equivalent forms. To provide a quantitative understanding of it, we derive analytical expressions for three common non-Maxwellian plasma distribution functions.  We show that there are undesirable features of this non-Maxwellianity measure including that it can diverge in various physical limits and elucidate the reason for the divergence. We then introduce a new kinetic entropy-based non-Maxwellianity measure based on the velocity-space kinetic entropy density, which has a meaningful physical interpretation and does not diverge. We use collisionless particle-in-cell simulations of two-dimensional anti-parallel magnetic reconnection to assess the kinetic entropy-based non-Maxwellianity measures. We show that regions of non-zero non-Maxwellianity are linked to kinetic processes occurring during magnetic reconnection.  We also show the simulated non-Maxwellianity agrees reasonably well with predictions for distributions resembling those calculated analytically. These results can be important for applications, as non-Maxwellianity can be used to identify regions of kinetic-scale physics or increased dissipation in plasmas.
\end{abstract}  


\section{Introduction}
\label{sec:intro2}

The conversion and dissipation of energy at small-scales in magnetized plasmas is a crucial aspect of many phenomena of importance to heliospheric and planetary science.  For example, heating of the solar corona to temperatures far greater than its surface is related to wave heating and turbulence [{\it e.g.}, \citep{Heyvaerts83,Nakariakov99,matthaeus1999a}] and magnetic reconnection underlying nano-flares [{\it e.g.,} \citep{Klimchuk06,Zank_etal_2018a}].  Local heating needs to occur in the turbulent solar wind to explain observed temperature profiles [{\it e.g.}, \citep{Gosling07,matthaeus1999b,Adhikari17,Adhikari20}].  Dynamics near and within Earth's bow shock plays an important role in setting the conditions of the plasma abutting Earth's magnetosphere [{\it e.g.,} \citep{Feldman82,Burgess12}].  Magnetic reconnection at both the dayside and the magnetotail is a crucial aspect of solar wind-magnetospheric coupling at Earth [{\it e.g.,} \citep{Levy64}] and Mercury [{\it e.g.,} \citep{Slavin09}] and dynamics in the magnetospheres of the outer planets [{\it e.g.,} \citep{Vasyliunas83,McAndrews08,Fuselier20}]. 

In collisional magnetized plasmas, the dissipation of energy at boundary layers in shocks, reconnecting current sheets, and intermittent current sheets in a turbulent medium is relatively well understood. However, in many settings of interest for heliospheric and planetary sciences, the plasma is weakly collisional, so collisions are too weak to influence the boundary layers. In such settings, the boundary layers are typically set by gyroscales of the constituent plasma species.  At these scales, the dynamics is dominated by kinetic physics, necessitating a kinetic description of the plasma.

Kinetic-scale dynamics historically was difficult to directly measure because it occurs on relatively short spatial and temporal scales. However, the measurement of kinetic features, including velocity distribution functions of the constituent plasma, is now achievable in kinetic simulations and {\it in situ} satellite observations. In particular, the Magnetospheric Multiscale (MMS) mission \citep{Burch16} can resolve both electron and ion kinetic scales spatially and temporally, providing an unprecedented and exquisite opportunity to learn about the kinetic physics underlying reconnection [{\it e.g.,} \citep{Burch16b,Torbert18}], turbulence [{\it e.g.,} \citep{Bandyopadhyay18}], and collisionless shocks [{\it e.g.} \citep{Gingell17,Goodrich18}].

There are many theoretical and analytical approaches to studying kinetic-scale energy conversion and dissipation. In this study, we focus on one underutilized quantity: kinetic entropy, {\it i.e.,} entropy defined fully within kinetic theory [{\it e.g.,} \citep{Liang19} and references therein]. The kinetic entropy is often written as being proportional to the phase space integral of $f\ln f$, where $f$ is the velocity distribution function of the plasma species. This is in contrast to the fluid entropy, related to $p/\rho^\gamma$ where $p$ is the pressure, $\rho$ is the mass density, and $\gamma$ is the ratio of specific heats, which is only valid for a plasma in local thermodynamic equilibrium (for which $f$ is a Maxwellian everywhere in space). Entropy has both desirable and undesirable properties. Its main desirable property is that it is uniquely related to irreversible dissipation in collisional systems \citep{Boltzmann77}, which potentially makes it a key quantity to identify regions where dissipation may be happening in systems of interest. Its main drawback is that the relation of entropy to dissipation is true for closed systems, but it is not clear that physical systems of interest can be construed as closed.

Consequently, while there have been numerous studies of the fluid form of entropy in heliospheric systems [see \citet{Liang19} for references], there are only a few studies investigating kinetic entropy. Observationally, there have been attempts to measure the kinetic entropy or use entropic measures with satellite observations of, for example, Earth's magnetotail plasma sheet \citep{Kaufmann09,Kaufmann11}, Earth's bow shock \citep{Parks12}, the near-Earth solar wind \citep{Weck15,Olivier19}, and auroral currents \citep{Osmane19}. It has also been used in a number of theoretical and numerical studies \citep{Montgomery70,Hsu74,Krommes94,Leubner04,Watanabe04,Howes06,Schekochihin09,Tatsuno09,Sarazin09,Nariyuki11,Nakata12,Loureiro13,Tenbarge13,Numata15,Pezzi16,Groselj17,Hesse17,Pezzi17,Eyink18,Gary18,Cerri18,Liang19,Pezzi19,Kawazura19,Du20}.

We focus on the work by \citet{Kaufmann09} in the present study. In their observational study of Earth's plasma sheet, they used the kinetic entropy per particle as a diagnostic in their observations.  One aspect of their study was to compare the kinetic entropy per particle to its fluid counterpart. The difference between the two at a given location and time gives a measure of how ``non-Maxwellian'' a plasma is, and therefore gives a measure of the importance of non-equilibrium kinetic effects.  This measure of non-Maxwellianity is not unique. Other non-Maxwellianity measures include the so-called $\epsilon$ parameter \citep{Greco12} and the so-called enstrophy \citep{Servidio17}.

Knowing and quantifying the non-Maxwellianity of a distribution function is potentially of great utility since dissipation is typically associated with the emergence of non-Maxwellian distribution functions and the collisional relaxation back towards Maxwellianity {\it e.g.,} \citet{Vaivads16,Valentini16,Matthaeus20}]. However, we are not aware of theoretical and/or computational studies which have put the entropy-based non-Maxwellianity measure on a firm footing.  In other words, what does it mean for a plasma to have a particular departure from the (equilibrium) fluid entropy? 

In this study, we provide a theoretical investigation of what we call the Kaufmann and Paterson non-Maxwellianity. We show that it has equivalent forms and provide a physical interpretation of these forms. Then, we perform an analytical calculation of it for three common closed-form non-Maxwellian distribution functions, namely two beams separated in velocity space, a bi-Maxwellian, and the distribution studied by \citet{egedal13} and colleagues that appears near magnetic reconnection sites. We then show that the Kaufmann and Paterson non-Maxwellianity has the undesirable property that it can diverge, and provide the underlying reason for this.  We then present a new non-Maxwellianity measure that does not diverge in the same limits.  The theoretical work is then tested with data from particle-in-cell simulations of magnetic reconnection. Links between the appearance of a non-zero non-Maxwellianity and the kinetic effects taking place during the reconnection process are made. Comparisons of the analytical non-Maxwellianity expressions are made with representative distributions that naturally arise in the simulations of reconnection, revealing good agreement.

This paper is organized as follows. In Section~\ref{sec:non_max_KP}, we review the definition of the Kaufmann and Paterson non-Maxwellianity. Section~\ref{sec:ent_nonmax_kp_theory} analyzes the quantity in general, and provides analytical expressions for three common distributions. Section~\ref{sec:ent_nonmaxsub1_liang} points out issues with the existing measure, explains the cause, presents a new non-Maxwellianity measure, and shows it eliminates the issues.  Section~\ref{sec:sim_setup} describes the setup of the particle-in-cell simulations. The simulation results and comparisons to the theory are shown in Section~\ref{sec:simResults}. Discussion and conclusions are provided in Section~\ref{sec:disc_conc}.  A comparison of the non-Maxwellianity measures discussed here with other quantities that have been used to identify kinetic-scale physics in weakly collisional plasmas is outside the scope of the present study, but is carried out in a companion study \citep{Pezzi20}. 

\section{Kaufmann and Paterson Kinetic Entropy-based Non-Maxwellianity}
\label{sec:non_max_KP}

Here, we review the kinetic entropy-based measure developed by \citet{Kaufmann09} to measure the non-Maxwellianity of an arbitrary given distribution function $f(\vec{r},\vec{v},t)$ as a function of position $\vec{r}$ and velocity $\vec{v}$ at a fixed time $t$.  (We henceforth suppress the $\vec{r}$ and $t$ dependence for simplicity.) First, one calculates the density $n = \int d^3v f(\vec{v})$, bulk velocity $\vec{u} = (1/n) \int d^3v \vec{v} f(\vec{v})$ and effective temperature $T~=~(m/3 n k_B) \int d^3v (\vec{v}-\vec{u})^2 f(\vec{v})$, where $k_B$ is Boltzmann's constant and $m$ is the mass of a particle. The Maxwellianized distribution $f_M(\vec{v})$ associated with $f(\vec{v})$ is defined as
\begin{equation}
f_{M}(\vec{v}) = n\left(\frac{m}{2 \pi k_B T}\right)^{3/2} e^{-m (\vec{v} - \vec{u})^2 / 2 k_B T}. \label{eq:maxwell}
\end{equation}
The local (continuous) kinetic entropy density $s$ [{\it e.g.}, Eq.~(3) in \citet{Liang19}] of the full distribution function $f(\vec{v})$ is
\begin{equation}
  s=-k_B\int d^3v f(\vec{v})\ln f(\vec{v}).
  \label{eq:entdens}
\end{equation}
The kinetic entropy density $s_M$ associated with the Maxwellianized distribution $f_M(\vec{v})$ is
\begin{equation}
  s_M = -k_B \int d^3v f_M(\vec{v}) \ln
  f_M(\vec{v}). \label{eq:smaxwell}
\end{equation}
Equation~(\ref{eq:smaxwell}) is analytically solvable using direct substitution of Eq.~(\ref{eq:maxwell}), giving
\begin{equation}
  s_M = \frac{3}{2}k_B n \left[1 + \ln \left( \frac{2\pi k_B T}{mn^{2/3}}\right) \right]. \label{eq:smaxwellcalc}
\end{equation}
This form motivated \citet{Kaufmann09} to define a non-Maxwellianity measure, which we denote $\bar{M}_{KP}$, as
\begin{equation}\label{eq:flnf_M}
  \bar{M}_{KP} = \frac{s_M - s}{(3/2)k_B n}.
\end{equation}
They chose to normalize to $(3/2)k_B n = c_v n$, where $c_v = (3/2)k_B$ is the specific heat per particle at constant volume for an ideal gas, so that $\bar{M}_{KP}$ is dimensionless.  They note, however, that the dimensions of $s$ and $s_M$ individually are not well-defined because they include a natural logarithm of the dimensional quantity $f(\vec{v})$. This is not an issue for differences in entropy density, which can be written as having a natural logarithm of a dimensionless quantity.  [See also Appendix B4 of \citet{Liang19}].

\section{Theory of the Kaufmann and Paterson Non-Maxwellianity}
\label{sec:ent_nonmax_kp_theory}

\subsection{Basic Properties of Kaufmann and Paterson Non-Maxwellianity}
\label{sec:basicpropmbarkp}

Here, we gather some basic properties about the Kaufmann and Paterson non-Maxwellianity measure $\bar{M}_{KP}$.  First, obviously, if $f(\vec{v})$ is Maxwellian, then $f_M(\vec{v}) = f(\vec{v})$ and $\bar{M}_{KP} = 0$. Second, it has long been known that $f_M(\vec{v})$ is the distribution with the maximum kinetic entropy for a fixed number of particles and total energy (in the absence of electromagnetic fields, net charge, and net current) [{\it e.g.}, \citep{Boltzmann77,Bellan08}].  Thus, $s_M$ is the maximum entropy density for a fixed number of particles and energy.  Therefore, if $\bar{M}_{KP} = 0$, then $f(\vec{v})$ is Maxwellian, and one expects $\bar{M}_{KP}$ to be strictly non-negative. For these reasons, $\bar{M}_{KP}$ is a good measure of non-Maxwellianity.

It is potentially a useful measure because it is a local measure which can identify regions with non-Maxwellian distributions. This is worthwhile to know because the rate of change of the local entropy density~$s$ is [{\it e.g.}, \citep{Eyink18}]
    \begin{equation}
  \frac{\partial s}{\partial t} + 
\nabla \cdot \vec{\mathcal{J}} = -k_B \int d^3v C[f(\vec{v})] [1 + \ln f(\vec{v})],\label{eq:entdensevol}
\end{equation}
where $\vec{\mathcal{J}} = -k_B \int d^3v \vec{v}f(\vec{v})\ln f(\vec{v}) $ is the entropy density flux and $C[f(\vec{v})]$ is the collision operator.  The collision operator for a single species typically vanishes if $f(\vec{v})$ is Maxwellian, so the degree of non-Maxwellianity can be related to dissipation through collisions [{\it e.g.}, \citep{Liang20}].  Caution is necessary, however,  because there are systems where dissipation occurs even if distributions are Maxwellian everywhere. One example is if the constituent species have come to equilibrium with themselves, but are at different temperatures than each other; there can be dissipation through inter-species collisions even though each distribution is Maxwellian [{\it e.g.}, \citep{Guo17,Groselj17,Parashar18,Arzamasskiy19,Kawazura19,Cerri19,Parashar19,Rowan19,Zhdankin19}].  A second example is at an infinitely thin shock; the non-Maxwellianity is zero everywhere in such a system, but there is dissipation and entropy production at the discontinuity.

The quantity $\bar{M}_{KP}$ is fluid-like, obtained from velocity space integrals of a function of the local distribution function. Thus, it should be able to be calculated using satellite, simulation, or laboratory experiment data not very differently than calculating moments of the distribution function such as density or temperature. 

Another important property of $\bar{M}_{KP}$ is that it is independent of density, as we now derive.  Dividing Eq.~(\ref{eq:entdens}) by $n$, then adding and subtracting $[f(\vec{v})/n]\ln n$ inside the integrand and simplifying gives
\begin{equation}\label{eq:fnlnfn2}
\frac{s}{n} = - k_B \int d^3v
\frac{f(\vec{v})}{n} \ln\left(
\frac{f(\vec{v})}{n}\right) - k_B \ln n.
\end{equation}
Using this result to directly calculate $\bar{M}_{KP} = (s_M-s)/(3/2)k_B n$ reveals that the $k_B \ln n$ term cancels because the densities associated
with $f(\vec{v})$ and $f_M(\vec{v})$ are the same, so
\begin{equation}
  \bar{M}_{KP} = \frac{2}{3} \left[ - \int d^3v \left(\frac{f_M(\vec{v})}{n}
  \right)\ln\left(\frac{f_M(\vec{v})}{n} \right) + \int
  d^3v \left(\frac{f(\vec{v})}{n} \right)\ln\left(\frac{f(\vec{v})}{n} \right) \right]. \label{eq:Mbar}
\end{equation} 
This shows that if one uses the convention where the distribution function is a probability density instead of a phase space density, {\it i.e.,} $f(\vec{v}) \rightarrow f(\vec{v})/n$, then the result for $\bar{M}_{KP}$ is unchanged. It also shows that $\bar{M}_{KP}$ has no explicit dependence on the plasma density $n$.

We note that $\bar{M}_{KP}$ contains similar information to the non-Maxwellianity parameter $\epsilon$ introduced by \citet{Greco12} and the enstrophy non-Maxwellianity $\Omega$ \citep{Servidio17}. In our notation, $\epsilon$ is
\begin{equation}  
\epsilon = \frac{1}{n}\sqrt{\int d^3v \left[f(\vec{v})-f_M(\vec{v})\right]^2}
 \label{eq:epsilon}
\end{equation}
and $\Omega = n^2 \epsilon^2$. The latter was simplified by expanding $f(\vec{v})$ in a Hermite expansion, which relates $\Omega$ to the Hermite spectrum of $f(\vec{v})$. In the limit that the departure from a Maxwellian is small, we can write $f(\vec{v}) = f_M(\vec{v}) + \delta f(\vec{v})$.  Doing an expansion of $\bar{M}_{KP}$ to second order in $\delta f(\vec{v})$ gives
\begin{equation}
    \bar{M}_{KP} \simeq \frac{1}{3n}\int d^3v \frac{[f(\vec{v}) - f_M(\vec{v})]^2}{f_M(\vec{v})},
\end{equation}
as is well-known in gyrokinetic theory [{\it e.g.,} \citep{Howes06,Groselj17,Cerri18,Kawazura19}]. This is quadratic in $\delta f(\vec{v})$, similar to $\epsilon$ and $\Omega$. Thus, one would expect $\epsilon$, $\Omega$, and $\bar{M}_{KP}$ to have similar structure in strongly collisional systems where the deviation from Maxwellian distributions is small. When deviations from Maxwellianity are large, the two measures likely are different.  These measures are compared with each other and other dissipation measures for weakly collisional systems in a companion study \citep{Pezzi20}.

This section provides some insight into the properties of $\bar{M}_{KP}$, but it does not address how to interpret what it means for the non-Maxwellianity to be a particular number.  The following sections introduce three examples where analytical values of $\bar{M}_{KP}$ are calculated for common non-Maxwellian distribution functions.

\subsection{Kaufmann and Paterson Non-Maxwellianity for Two Beams}
\label{sec:ent_nonmaxsub1_beams}

We calculate $\bar{M}_{KP}$ analytically for a two-population plasma that are each Maxwellian but drift parallel or anti-parallel to each other, and we require that the relative velocity of the beams is large enough that the overlap between the two populations in velocity space is negligible. A condition for this is derived below. The distribution function $f_{beam}(\vec{v})$ for such a system is given by
\begin{equation}
f_{beam}(\vec{v}) = n_1\left(\frac{m}{2 \pi k_B T_1}\right)^{3/2}
    e^{-m (\vec{v} - u_{z1}{\bf \hat{z}})^2 / 
    2k_B T_1} + n_2\left(\frac{m}{2 \pi k_B T_2}\right)^{3/2}  
    e^{-m (\vec{v} - u_{z2} {\bf \hat{z}})^2 / 2k_B T_2},
  \label{eq:fbeam}
\end{equation}
where $n_1$ and $n_2$ are the densities of the two beams, $u_{z1}$ and $u_{z2}$ are the bulk velocities of the two beams, assumed parallel or anti-parallel, and $T_1$ and $T_2$ are the temperatures of the two individual beams. By taking moments, it is straight-forward to show that the density, bulk flow, and effective temperature are
\begin{eqnarray}
    n & = & n_1 + n_2, \\
    u_{z} & = & \frac{n_1 u_{z1} + n_2 u_{z2}}{n_1 + n_2}, \\
    T_{beam} & = & \frac{mn_1n_2}{3k_B (n_1+n_2)^2}(u_{z1}-u_{z2})^2 + \frac{n_1 T_1 + n_2 T_2}{n_1 + n_2}. \label{eq:tmbeam}
\end{eqnarray}
These bulk properties are valid independent of whether the two populations overlap in velocity space.  The kinetic entropy density, however, is not exactly solvable unless the overlap between the two distributions is negligible, which occurs when the first term in Eq.~(\ref{eq:tmbeam}) dominates the second term. In that limit, the kinetic entropy density $s_{beam}$ from Eq.~(\ref{eq:entdens}) is just the sum of the kinetic entropies of the individual beams,
\begin{equation}
    s_{beam} \simeq \frac{3}{2}k_B (n_1 + n_2) + \frac{3}{2} k_B \left[ n_1 \ln \left( \frac{2\pi k_B T_1}{m n_1^{2/3}} \right) + n_2 \ln \left( \frac{2\pi k_B T_2}{m n_2^{2/3}} \right) \right].
\end{equation}
Eq.~(\ref{eq:flnf_M}) and (\ref{eq:smaxwellcalc}) give an associated non-Maxwellianity of
\begin{equation}
\bar{M}_{KP,beam} \simeq \ln \left( \frac{T_{beam}/n^{2/3}}{(T_1/n_1^{2/3})^{n_1/n} (T_2/n_2^{2/3})^{n_2/n}} \right). \label{eq:mbarkpbeam}
\end{equation}

As a special case, if the beams are identical plasmas ($n_1 = n_2$ and $T_1 = T_2$) and they are counter-propagating ($u_{z1} = - u_{z2}$), then
\begin{equation}
\bar{M}_{KP,beam} \simeq \ln \left( \frac{T_{beam}}{2^{2/3} T_1} \right) \simeq \ln \left( \frac{m u_{z1}^2 / 3 + k_B T_1}{2^{2/3} k_B T_1} \right). \label{eq:mbarkpbeamexample}
\end{equation}
Letting $u_{z1}^2 = {\cal M}^2 k_B T_1 / m$ with ${\cal M} \gg 1$, where ${\cal M}$ is an effective Mach number of the flow (leaving out a factor of the ratio of specific heats $\gamma$), then the Kaufmann and Paterson non-Maxwellianity for this distribution is $\bar{M}_{KP,beam} \simeq \ln [({\cal M}^2/3 + 1)/2^{2/3}]$. 

\subsection{Kaufmann and Paterson Non-Maxwellianity for Bi-Maxwellian Distributions}
\label{sec:ent_nonmaxsub1_bimax}

A bi-Maxwellian distribution function $f_{biM}(\vec{v})$ is defined as
\begin{equation}
f_{biM}(\vec{v}) = n\left(\frac{m}{2 \pi k_B T_\perp}\right)\left(\frac{m}{2 \pi k_B T_\parallel}\right)^{1/2}  
    e^{-m (\vec{v} - \vec{u})_\perp^2 / 
    2k_B T_\perp} e^{-m (\vec{v} - \vec{u})_\parallel^2 / 2
  k_B T_\parallel},
  \label{eq:bimaxwell}
\end{equation}
where the $\perp$ and $\parallel$ subscripts allow for anisotropic velocities and temperatures, typically relative to the direction of a magnetic field. Straight-forward calculation of the associated kinetic entropy density from Eq.~(\ref{eq:entdens}) gives
\begin{equation}
  s_{biM} = \frac{3}{2}k_B n \left[1 + \ln \left( \frac{2\pi k_B T_\perp^{2/3}T_\parallel^{1/3}}{mn^{2/3}}\right) \right], \label{eq:sbimaxwellcalc}
\end{equation}
and Eq.~(\ref{eq:flnf_M}) gives an associated non-Maxwellianity of
\begin{equation}
\bar{M}_{KP,biM} = \ln \left( \frac{T}{T_\perp^{2/3}T_\parallel^{1/3}} \right) = \ln \left[ \frac{2}{3} \left( \frac{T_\perp}{T_\parallel}\right)^{1/3} + \frac{1}{3}\left(\frac{T_\parallel}{T_\perp}\right)^{2/3} \right], \label{eq:mbarkpbimax}
\end{equation}
where the second form eliminates the effective temperature using $T = (2/3) T_\perp + (1/3) T_\parallel$.

\begin{figure}
  \includegraphics[width=3.4in]{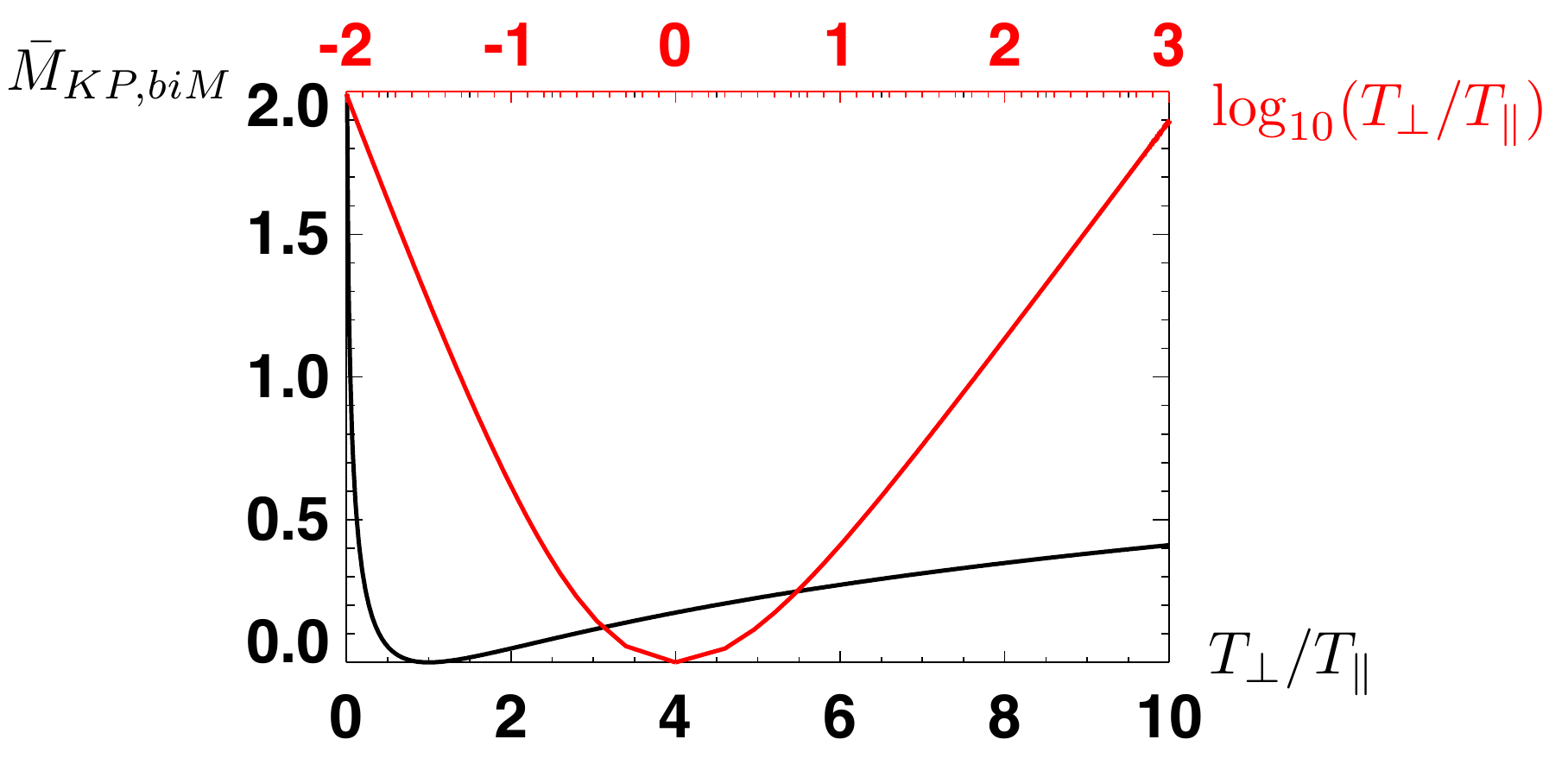}
  \centering
\caption{Plot of the Kaufmann and Paterson non-Maxwellianity $\bar{M}_{KP,biM}$ for a bi-Maxwellian distribution function $f_{biM}(\vec{v})$ as a function of the ratio of perpendicular to parallel temperature $T_\perp/T_\parallel$. The black line uses a linear horizontal scale on the bottom axis, and the red line uses a logarithmic horizontal scale on the top axis over a wider range of $T_\perp/T_\parallel$ to show that it diverges for small and large $T_\perp/T_\parallel$.} \label{fig:nonmax_bimax}
\end{figure}
A plot of $\bar{M}_{KP,biM}$ as a function of $T_\perp/T_\parallel$ is given in black on a linear scale in Fig.~\ref{fig:nonmax_bimax}.  This helps give perspective on values of the Kaufmann and Paterson non-Maxwellianity measure for a bi-Maxwellian distribution function.  In particular, $\bar{M}_{KP,bim} = 0$ for a Maxwellian plasma ($T_\perp / T_\parallel = 1$), as expected.  For example values, $\bar{M}_{KP,biM} \simeq 0.17$ for $T_\perp/T_\parallel = 4$ and $\bar{M}_{KP,biM} \simeq 0.23$ for $T_\perp/T_\parallel = 1/4$.  

Interestingly, Eq.~(\ref{eq:mbarkpbimax}) reveals that $\bar{M}_{KP,biM}$ diverges to infinity as $T_\perp/T_\parallel$ goes to either zero or infinity.  The red line in Fig.~\ref{fig:nonmax_bimax} uses a logarithmic horizontal scale over a broader range of $T_\perp / T_\parallel$ to motivate this.  The reason for the divergence is discussed in Section~\ref{sec:ent_nonmaxsub1_liang}.

\subsection{Kaufmann and Paterson Non-Maxwellianity for Egedal Distributions}
\label{sec:ent_nonmaxsub1_egedal}

During magnetic reconnection, magnetic fields in the upstream region bend as they approach the reconnection site. A magnetic field-aligned electric field accelerates electrons into this region, leading to a population of electrons that gets trapped in the mirror field \citep{egedal13}. The electron velocity distribution functions in these regions are elongated in the direction parallel to the magnetic field, leading to a gyrotropic distribution. The distribution is a double adiabatic and reversible solution to the electron drift kinetic equation obtained in the the limit of short electron transit/bounce time \citep{egedal13}. Here, we call it an Egedal distribution $f_{Eg}(\vec{v})$, and it is given by
\begin{equation}
f_{Eg}(\vec{v}) = 
\begin{cases}
  n_{\infty}\left(\frac{2 \pi k_B T_{\infty}}{m}\right)^{-3 / 2} e^{-\frac{m v_\perp^{2} B_{\infty}}{2 k_B T_{\infty} B}} & \text{trapped} \\
   n_{\infty}\left(\frac{2 \pi k_B T_{\infty}}{m}\right)^{-3 / 2} e^{-\frac{m (v_\perp^{2} + v_\parallel^{2})}{2 k_B T_{\infty}}} e^{\frac{e \phi_\parallel}{k_B T_{\infty}}} & \text{passing} 
  \label{eq:Egedal} 
\end{cases}\
\end{equation}
where $n_\infty$, $T_\infty$ and $B_\infty$ are the density, temperature and magnetic field strength far upstream, $B$ is the local magnetic field strength, $\phi_\parallel$ is the parallel acceleration potential, and $v_\perp$ and $v_\parallel$ are the speeds perpendicular and parallel to the magnetic field. The trapped/passing boundary is given by
\begin{equation}
    \frac{1}{2} m\left(v_\parallel^{2}+v_\perp^{2}\right)-e \phi_{\parallel}-\frac{1}{2} \frac{m v_\perp^{2}}{B} B_{\infty}=0.
    \label{passingtrappedboundary}
\end{equation}

Calculating the local number density $n = \int d^3v f_{Eg}(\vec{v})$ for this distribution gives \citep{Le09}
\begin{equation}
\frac{n}{n_{\infty}} = 2b \sqrt{\frac{\Phi}{\pi}} + \text {erfcx} \left( \sqrt{\Phi} \right) - (1- b)^{3/2} \text {erfcx} \left(\sqrt{\frac {\Phi}{1-b}} \right) 
\label{I4def}    
\end{equation}
where erfcx($x$) = $e^{x^2}$ erfc($x$) = $e^{x^2}$ [1 - erf($x$)] is the scaled complementary error function, erfc($x$) = $(2/\sqrt{\pi})\int_x^\infty e^{-z^2} dz$, $b = B/B_\infty$, and $\Phi = e\phi_\parallel / k_B T_\infty$. Note, in the limit of $\Phi \rightarrow 0$ and $b \rightarrow 1$, the trapped/passing boundary from Eq.~(\ref{passingtrappedboundary}) reduces to a point at $v_\parallel = 0$, and the distribution function $f_{Eg}(\vec{v})$ reduces to a Maxwellian, so the Maxwellian results should be recovered.  Since erfcx(0) = 1, we recover $n=n_{\infty}$ in this limit, as expected.

The kinetic entropy density $s_{Eg}$ for an Egedal distribution follows from direct application of Eq.~(\ref{eq:entdens}). A lengthy calculation gives
\begin{equation}
s_{Eg} = \frac{3}{2} k_B n \left[ \frac{n_\infty G}{n} + \ln \left( \frac{2 \pi k_B T_\infty}{m n_\infty^{2/3}} \right) \right],
\label{eq:sEgedalcalc}
\end{equation}
where 
\begin{equation}
    G = 2b \sqrt{\frac{\Phi}{\pi}} +\left(1 - \frac{2\Phi}{3}\right) \text {erfcx} \left( \sqrt{\Phi} \right) - \sqrt{1-b} \left(1 - b -\frac{2\Phi}{3}\right)   \text {erfcx}\left(\sqrt{\frac{\Phi}{1-b}}\right).
\label{L567def}
\end{equation}
As a check, in the $\Phi \rightarrow 0, b \rightarrow 1$ limit, $G \rightarrow 1$, so Eq.~(\ref{eq:sEgedalcalc}) reduces to Eq.~(\ref{eq:smaxwellcalc}), as expected.  We also note that, since erfcx$(x) \rightarrow 1/(x\sqrt{\pi})$ asymptotically in the $x \rightarrow \infty$ limit, $s_{Eg}$ diverges as $\Phi \rightarrow \infty$.

To calculate $\bar{M}_{KP,Eg}$ for Egedal distributions from Eq.~(\ref{eq:flnf_M}), one needs the effective temperature $T_{Eg}$ for Egedal distributions to get the entropy density of the Maxwellianized distribution. The parallel temperature $T_{\parallel,Eg} = [m/(n k_B)] \int d^3v (v_\parallel-u_\parallel)^2 f(\vec{v})$ and perpendicular temperature $T_{\perp,Eg} = [m/(2n k_B)] \int d^3v (\vec{v}_\perp-\vec{u}_\perp)^2 f(\vec{v})$, following lengthy calculations, are
\begin{eqnarray}
  T_{\parallel,Eg} & = & \frac{n_\infty T_{\infty}}{n} \left[ \text {erfcx} \left( \sqrt{\Phi} \right) + 2b\left(2-b + \frac{2\Phi}{3}\right)\sqrt{\frac{\Phi}{\pi}}  \right. \nonumber \\ & & \left. -\left(1-b\right)^{5/2} \text {erfcx}\left( \sqrt{\frac{\Phi}{1-b}}\right)\right],
\label{Tparalocaldef} \\ 
     T_{\perp,Eg} & = & \frac{n_\infty T_{\infty}}{n} \left\{\text {erfcx} \left(\sqrt{\Phi} \right) + b (3b - 1) \sqrt{\frac{\Phi}{\pi}} \right. \nonumber \\
         & & \left.+ (1-b)^{3/2} \left[\frac{\Phi b}{1-b} - \left( \frac{3b}{2} + 1\right) \right] \text {erfcx}\left(\sqrt{\frac{\Phi}{1-b}}\right) \right\},
\label{Tperplocaldef} \\
T_{Eg} & = & \frac{2}{3} T_{\perp,Eg} +\frac{1}{3} T_{\parallel,Eg}.
\label{TlocalEg}
\end{eqnarray}
As a check, $T_{\parallel,Eg}$, $T_{\perp,Eg}$, and $T_{Eg}$ all go to $T_\infty$ in the $\Phi \rightarrow 0, b \rightarrow 1$ limit, as expected.  Then, $s_M$ is calculated from Eq.~(\ref{eq:smaxwellcalc}) and using the result with Eqs.~(\ref{eq:sEgedalcalc}) and (\ref{eq:flnf_M}), the closed-form non-Maxwellianity $\bar{M}_{KP,Eg}$ for Egedal distribution functions is
\begin{equation}
\bar{M}_{KP,Eg} = 1 - \frac{n_\infty G}{n} + \ln \left(\frac{T_{Eg}/n^{2/3}}{T_\infty/n_\infty^{2/3}} \right). \label{eq:mbaregedal}
\end{equation}

\begin{figure}
\centering
    \includegraphics[width=0.9\linewidth]{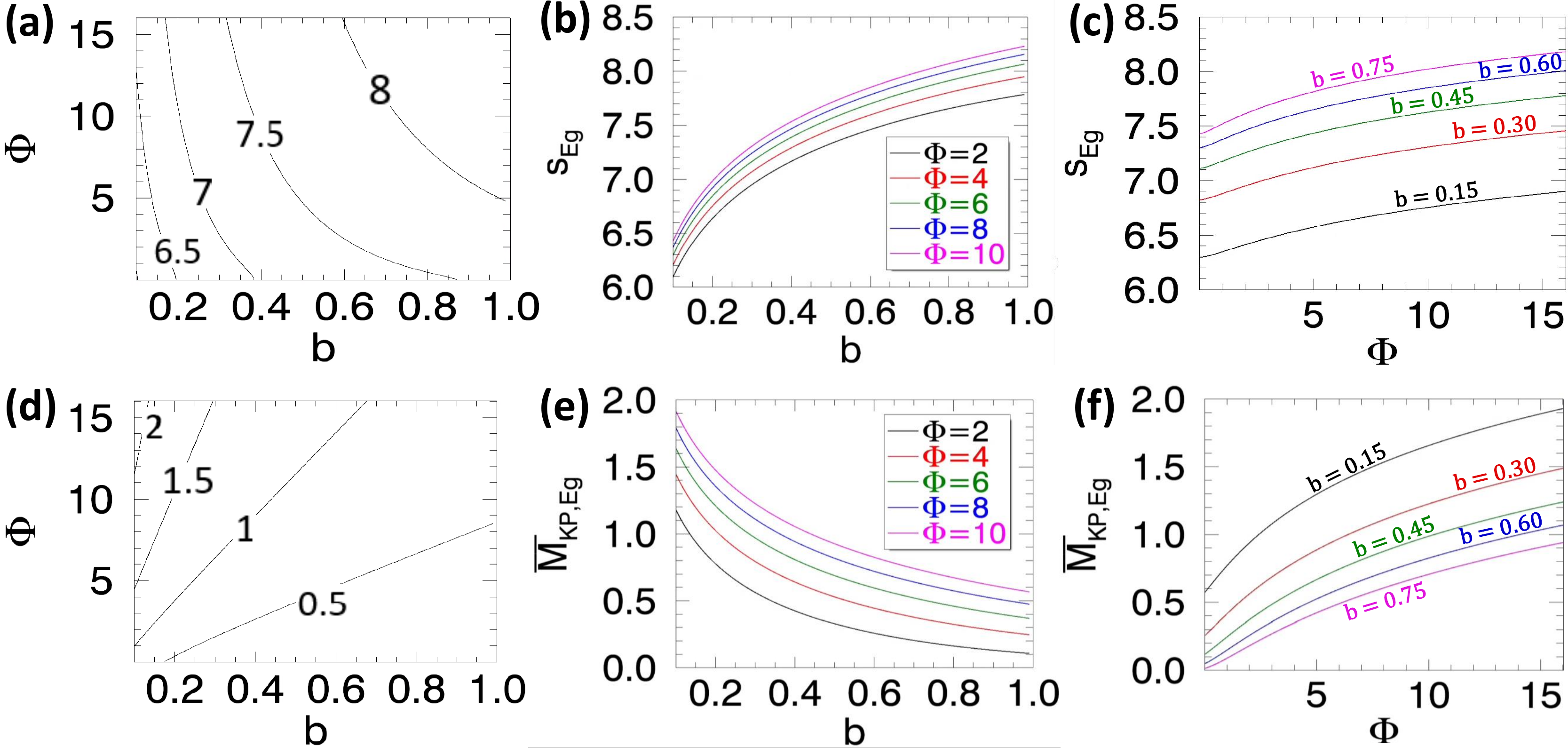}
    \caption{Kinetic entropy density $s_{Eg}$ and Kaufmann and Paterson non-Maxwellianity $\bar{M}_{KP,Eg}$ for an Egedal distribution function from Eq.~(\ref{eq:sEgedalcalc}) and Eq.~(\ref{eq:mbaregedal}) assuming $n/n_{\infty}$=0.805 and $T_\infty~=~0.08~B_\infty^2 / 4 \pi k_B n_\infty$. (a) and (d) are contour plots of $s_{Eg}$ and $\bar{M}_{KP,Eg}$, respectively, as a function of $b = B/B_\infty$ and $\Phi = e\phi_\parallel / k_B T_\infty$. (b) and (e) are cuts of these as a function of $b$ for five representative values of $\Phi$.  (c) and (f) are cuts of these as a function of $\Phi$ for five representative values of $b$.}
    \label{fig:S_Eg_MBar_KPEgPlots}
\end{figure}

For reference, plots of kinetic entropy density $s_{Eg}$ and Kaufmann and Paterson non-Maxwellianity $\bar{M}_{KP,Eg}$ for an Egedal distribution are in Fig.~\ref{fig:S_Eg_MBar_KPEgPlots}, using a density of $n/n_{\infty}$=0.805 and $T_\infty~=~0.08~B_\infty^2 / 4 \pi k_B n_\infty$. Panels (a) and (d) are contour plots of $s_{Eg}$ and $\bar{M}_{KP,Eg}$, respectively, as a function of $b$ and $\Phi$. The former is normalized to $k_B n_\infty$. Panels (b) and (e) give cuts as a function of $b$ at $\Phi = 2, 4, 6, 8,$ and 10.  Panels (c) and (f) give cuts as a function of $\Phi$ at $b = 0.15, 0.30, 0.45, 0.60$ and 0.75.  The plots show that the non-Maxwellianity increases as $\Phi$ increases, which makes sense physically because this increases the temperature anisotropy leading to an increase in $\bar{M}_{KP}$, similar to bi-Maxwellian distributions in the previous section.

\begin{figure}
\centering
    \includegraphics[width=0.9\linewidth]{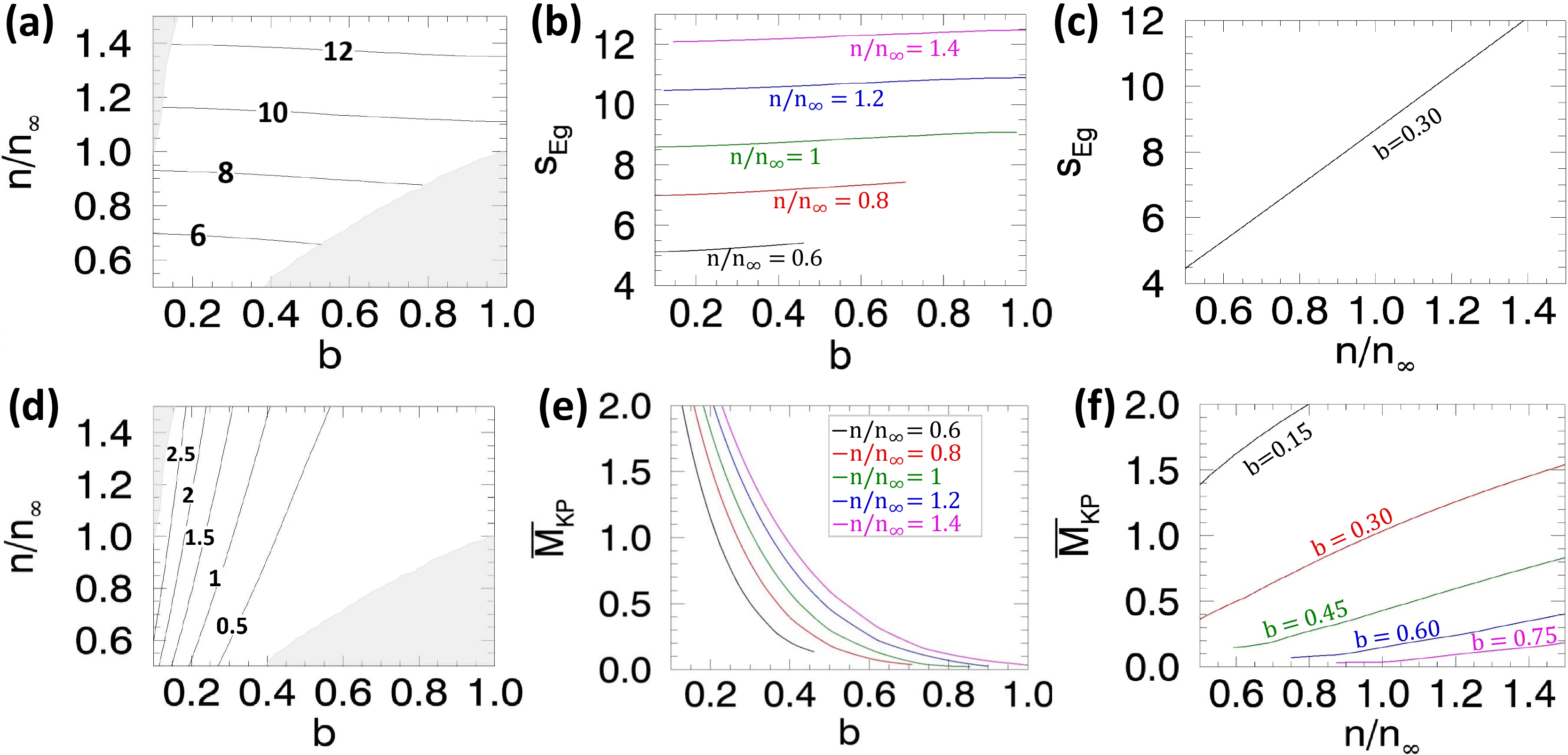}
\caption{Analogous to Fig.~\ref{fig:S_Eg_MBar_KPEgPlots}, except plotted as a function of $n/n_\infty$ and $b = B / B_\infty$ upon inversion of Eq.~(\ref{I4def}). The shaded regions in (a) and (d) correspond to parameters for which the inversion gives values of $\Phi$ below 0 or above 80, and are removed from the plot.}
    \label{fig:S_Eg_MBar_KPEgPlots_noPhi}
\end{figure}

Following \citet{Le09}, it is typically more useful to eliminate $\Phi$ in favor of $n/n_\infty$ and $b$ by numerically inverting Eq.~(\ref{I4def}). The result is then in terms of quantities more easily found in observations and simulations. Plots analogous to Fig.~\ref{fig:S_Eg_MBar_KPEgPlots} but as a function of $n/n_\infty$ and $b$ are in Fig.~\ref{fig:S_Eg_MBar_KPEgPlots_noPhi}. Panels (a) and (d) are contour plots of $s_{Eg}$ and $\bar{M}_{KP,Eg}$, respectively.  Panels (b) and (e) give cuts as a function of $b$ for $n/n_\infty = 0.6, 0.8, 1.0, 1.2,$ and 1.4. Panels (c) and (f) give cuts as a function of $n/n_\infty$ for fixed $b$; only $b = 0.3$ is shown in (c) since the dependence on $b$ is weak, while (f) shows cuts for $b = 0.15, 0.30, 0.45, 0.60$ and 0.75. Note that numerically inverting Eq.~(\ref{I4def}) gives negative $\Phi$ or extremely high $\Phi$ ($\geq 80$) for some values of $n/n_\infty$ and $b$. Such values are eliminated from the plots and are denoted by shaded gray regions in Fig.~\ref{fig:S_Eg_MBar_KPEgPlots_noPhi}(a) and (d).

The past three subsections provide exact solutions for the non-Maxwellianity measure of analytic forms of three common non-Maxwellian velocity distribution functions. These are potentially useful to quantify the non-Maxwellianity of self-consistently generated distribution functions in physical systems, such as those undergoing reconnection, turbulence, or shocks in magnetized plasmas. In self-consistent plasmas, the distributions undoubtedly are not exactly given by the expressions analyzed here, but should provide a reasonable approximation in some settings.  A test of this will be carried out for the reconnection simulations discussed in Secs.~\ref{sec:sim_setup} and \ref{sec:resultsdist}.

\section{A New Non-Maxwellianity Measure}
\label{sec:ent_nonmaxsub1_liang}

\subsection{Why the Kaufmann and Paterson  non-Maxwellianity Diverges}
\label{sec:theproblem1}

Desirable properties of the Kaufmann and Paterson non-Maxwellianity measure are that it is dimensionless, non-negative, and vanishes when the distribution function is Maxwellian.  An undesirable property of $\bar{M}_{KP}$ is that there is no upper bound, as shown in the previous section.  This makes it difficult to interpret what it means for the non-Maxwellianity to have a particular value. It would be preferable to have a normalized non-Maxwellianity measure that remains finite to facilitate its interpretation.

To develop such a measure, we must elucidate the cause of the divergence of $\bar{M}_{KP}$.  We see that it is not an issue with the definition of the non-Maxwellianity itself, but rather a fundamental issue with the kinetic theory description. Indeed, the entropy density $s_M$ of a Maxwellian, from Eq.~(\ref{eq:smaxwellcalc}), diverges for either $T \rightarrow 0$ or $T \rightarrow \infty$.

The problem arises as soon as one approximates the entropy density by the velocity space integral in Eq.~(\ref{eq:entdens}) instead of the combinatorial Boltzmann entropy related to the logarithm of the number of different microstates to produce a given macrostate. The cause of the problem is the coarse graining that is necessary to formulate the kinetic theory description.  As reviewed, for example, in \citet{Liang19}, in order to define kinetic entropy, or even a distribution function itself, one needs to break phase space into cells of hypervolume $\Delta^3 r \Delta^3 v$, where $\Delta^3 r$ is the spatial volume and $\Delta^3 v$ is the velocity space volume.  The size of these cells is restricted -- they cannot be too large where they do not resolve relevant structures in phase space, and they cannot be too small or the number of particles becomes too small for a statistical description.  This provides insight to why the kinetic entropy diverges. As $T \rightarrow 0$, velocity space structures become strongly peaked (mathematically, they approach a $\delta$-function), and a finite sized grid no longer resolves the structure.  As $T \rightarrow \infty$ for a fixed velocity-space grid, the number of particles in each phase space cell decreases, and the statistical description of the particles breaks down. 

These issues lead to unphysical results for the kinetic entropy using the standard definition from Eq.~(\ref{eq:entdens}) because the kinetic entropy should not diverge in these limits. To see this, note that evaluating Eq.~(\ref{eq:entdens}) for a $\delta$-function distribution function gives an $s$ that diverges. However, this divergence is specious. To justify this statement, we go back to the original combinatorial form of kinetic entropy given by Boltzmann in which $S = k_B \ln \Omega$, where $\Omega$ is the number of microstates corresponding to a given macrostate [see, {\it e.g.}, Appendix A1 of \citet{Liang19}]. The statistical interpretation of kinetic entropy is related to the number of ways to exchange the positions and velocities of particles in the system. For a $\delta$-function distribution, all particles are in a single cell in velocity space, so there is only one microstate for this macrostate. The combinatorial kinetic entropy is therefore $S = 0$.  Thus, Eq.~(\ref{eq:entdens}) giving $s = \infty$ is completely wrong in this limit.

Consequently, the divergence of the kinetic entropy is caused by a fundamental breakdown of kinetic theory as distributions get too broad or too peaked. The core reason for the problem is that in applications, the kinetic entropy, and indeed the distribution function itself, are only defined in a course-grained phase space, and is fundamentally an explicit function of the phase space grid scale.  The formulation of kinetic entropy producing Eq.~(\ref{eq:entdens}) does not capture this dependence, and this must be addressed to produce a non-Maxwellianity measure that is capable of being interpreted physically.

\subsection{The Non-Locality of the Kaufmann and Paterson non-Maxwellianity}
\label{sec:theproblem2}

A second fundamental issue with the Kaufmann and Paterson non-Maxwellianity measure is that one desires it to be a local measure. However, the kinetic entropy density $s$ contains information, in the combinatorial sense, about exchanging particles at different positions. Thus, using $s$ makes the non-Maxwellianity non-local in position space.  It is preferable to have the non-Maxwellianity, in the combinatorial sense, to locally describe only particles at a particular location being exchanged in velocity space.  It has been shown \citep{Mouhot11,Liang19} that the full kinetic entropy can be decomposed into a sum of a velocity space kinetic entropy and a position space kinetic entropy.  The velocity space kinetic entropy density $s_{{\rm velocity}}$ is
\begin{equation}
    s_{{\rm velocity}} = k_B \left[n \ln \left( \frac{n}{\Delta^3 v} \right) -\int d^3v f(\vec{v}) \ln f(\vec{v}) \right] = k_B n \ln \left( \frac{n}{\Delta^3 v} \right) + s.
\end{equation}
We argue that this kinetic entropy is more appropriate for defining a local measure of non-Maxwellianity.

Interestingly, a non-Maxwellianity analogous to the Kaufmann and Paterson definition but using velocity space kinetic entropy density is exactly equivalent to $\bar{M}_{KP}$, {\it i.e.,}
\begin{equation}
\bar{M}_{KP} = \frac{s_{\text{velocity},M} -
s_{\text{velocity}}}{(3/2)k_B n}.
\end{equation}
To see this, note that the density $n$ of the raw distribution $f(\vec{v})$ and the density of the Maxwellian distribution $f_M(\vec{v})$ are the same by definition, so the additional term $k_B n \ln (n/\Delta^3v)$ is the same for the raw and Maxwellianized distributions.  Thus, that term drops out of $s_{{\rm velocity},M} - s_{{\rm velocity}}$, and the resultant non-Maxwellianity is identical to $\bar{M}_{KP}$.  Thus, as far as the non-Maxwellianity measure is concerned, the Kaufmann and Paterson definition does give the desired result; the conclusion of this section is that a reinterpretation in terms of velocity space kinetic entropy density is desirable.

\subsection{A New Non-Maxwellianity Measure}
\label{sec:defmbar}

To address the issues discussed in the previous two sections, we propose a definition of the following normalized, non-divergent, non-Maxwellianity measure, which we denote $\bar{M}$:
\begin{equation}
  \bar{M} = \frac{s_{{\rm velocity},M} - s_{\rm velocity}}{s_{{\rm velocity},M}}. \label{eq:newmbar}
\end{equation}
This can be written equivalently as
\begin{equation}
  \bar{M} = \frac{-\int d^3v f_M(\vec{v}) \ln
  f_M(\vec{v}) + \int d^3v f(\vec{v})\ln f(\vec{v})}{n \ln \left( \frac{n}{\Delta^3 v} \right) - \int d^3v f_M(\vec{v}) \ln f_M(\vec{v})} = \frac{s_M - s}{s_M + k_B n \ln \left( \frac{n}{\Delta^3 v} \right)}. \label{eq:newmbar3}
\end{equation}
It can also be written in terms of $\bar{M}_{KP}$ by using Eq.~(\ref{eq:smaxwellcalc}) for $s_M$ in the denominator, resulting in
\begin{equation}
  \bar{M} = \frac{\bar{M}_{KP}}{1 + \ln \left( \frac{2\pi k_B T}{m(\Delta^3 v)^{2/3}} \right)}. \label{eq:newmbar2}
\end{equation}

The $\bar{M}$ measure retains the desirable properties of $\bar{M}_{KP}$.  First, it remains dimensionless; this is because a simple calculation confirms that $s_{{\rm velocity}}$ has appropriate dimensions of entropy per unit volume, unlike $s$ for which the dimensions are not well defined \citep{Kaufmann09,Liang19}. Second, as with $\bar{M}_{KP}$, we have $\bar{M} = 0$ if and only if the distribution function $f(\vec{v})$ is Maxwellian. Third, $\bar{M}$ is non-negative, provided that $\Delta^3v$ is appropriately chosen.  Fourth, from Eq.~(\ref{eq:newmbar2}), $\bar{M}$ is the same whether using distribution functions as phase space densities or probability densities, as was the case for $\bar{M}_{KP}$.

The $\bar{M}$ measure also addresses the issues in the previous two subsections.  It is a measure of the non-Maxwellianity that is local in position space since it is based on the velocity space kinetic entropy density. Also, $\bar{M}$ retains an explicit dependence on the velocity space grid scale, which may seem undesirable but as argued in Section~\ref{sec:theproblem1} is actually a fundamental requirement of the formulation of entropy and distribution functions within kinetic theory. It is the presence of $\Delta^3v$ that allows one to regularize the divergence that arises in $s$ and $s_M$, which ensures $\bar{M}$ is finite for any temperature provided an appropriate velocity space grid scale is chosen to properly resolve both $f(\vec{v})$ and $f_M(\vec{v})$.  

To see this, we evaluate $\bar{M}_{biM}$ for a bi-Maxwellian distribution.  From Eq.~(\ref{eq:newmbar2}) and Eq.~(\ref{eq:mbarkpbimax}), we get
\begin{equation}
  \bar{M}_{biM} = \frac{\ln \left[ \frac{2}{3} \left( \frac{T_\perp}{T_\parallel}\right)^{1/3} + \frac{1}{3}\left(\frac{T_\parallel}{T_\perp}\right)^{2/3} \right]}{1 + \ln \left( \frac{2\pi k_B [(2/3)T_\perp+(1/3)T_\parallel]}{m(\Delta^3 v)^{2/3}} \right)}, \label{eq:mbar_bim}
\end{equation}
where we use $T = (2/3)T_\perp + (1/3)T_\parallel$ in the denominator.  This expression is general, for any temperatures and velocity space grid scale.  To be specific, we consider a velocity space grid scale that is comparable to the thermal speed. This has been confirmed numerically in particle-in-cell simulations to be a good choice for the grid scale \citep{Liang19,Liang20}. Thus, we let the velocity space grid scale in the parallel and perpendicular directions be $\Delta v_\parallel = \alpha_\parallel (2 k_B T_\parallel / m)^{1/2}$ and $\Delta v_\perp = \alpha_\perp (2 k_B T_\perp / m)^{1/2}$, where $\alpha_\parallel$ and $\alpha_\perp$ are temperature-independent constants. With $\Delta^3v = \alpha_\perp^2 \alpha_\parallel (2 k_B T_\perp/m)(2k_BT_\parallel/m)^{1/2}$, Eq.~(\ref{eq:mbar_bim}) becomes
\begin{equation}
  \bar{M}_{biM} = \frac{\ln \left[ \frac{2}{3} \left( \frac{T_\perp}{T_\parallel}\right)^{1/3} + \frac{1}{3}\left(\frac{T_\parallel}{T_\perp}\right)^{2/3} \right]}{1 + \ln \left( \frac{\pi}{\alpha_\perp^{4/3} \alpha_\parallel^{2/3}} \right) + \ln \left[  \frac{2}{3} \left( \frac{T_\perp}{T_\parallel}\right)^{1/3} + \frac{1}{3}\left(\frac{T_\parallel}{T_\perp}\right)^{2/3}\right]}. \label{eq:mbar_bim_grid1}
\end{equation}
We immediately see from this form that 
\begin{equation}
  \lim_{T_\perp \rightarrow 0} \bar{M}_{biM} = 
  \lim_{T_\parallel \rightarrow 0} \bar{M}_{biM} = \lim_{T_\perp \rightarrow \infty} \bar{M}_{biM} = \lim_{T_\parallel \rightarrow \infty} \bar{M}_{biM} = 1, \label{eq:nonmax_bim_limits}
\end{equation}
so the non-Maxwellianity is regularized to a maximum of 1, independent of $\alpha_\perp$ and $\alpha_\parallel$.  This suggests the new non-Maxwellianity measure does have an interpretation as a fraction of the largest possible non-Maxwellianity, at least for a bi-Maxwellian distribution.
\begin{figure}
  \includegraphics[width=3.4in]{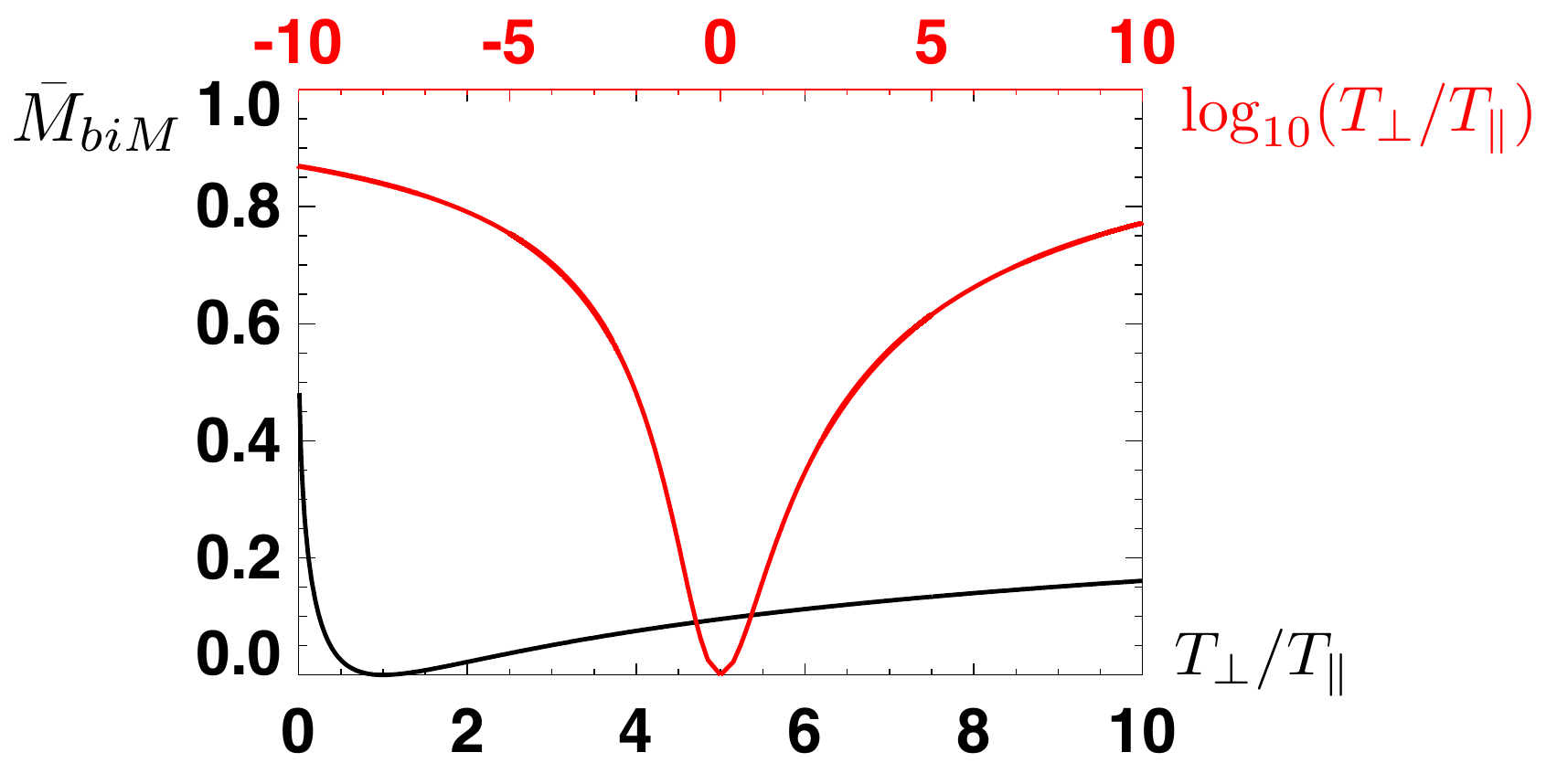}
  \centering
\caption{Plot of the non-Maxwellianity $\bar{M}_{biM}$ for a bi-Maxwellian distribution function as a function of $T_\perp/T_\parallel$ for $\alpha_\parallel = \alpha_\perp = 1$. The black line uses a linear horizontal scale on the bottom axis. The red line uses a logarithmic horizontal scale over a wider range of $T_\perp/T_\parallel$, and motivates that $\bar{M}_{biM}$ is finite even for small or large $T_\perp/T_\parallel$.} \label{fig:nonmax_bimax_hl}
\end{figure}
A plot of $\bar{M}_{biM}$ as a function of $T_\perp/T_\parallel$ is given in Fig.~\ref{fig:nonmax_bimax_hl} for $\alpha_\perp = \alpha_\parallel = 1$.  The black line is for a linear horizontal scale. Example values for this choice of the $\alpha$s are $\bar{M}_{biM} \simeq 0.075$ for $T_\perp/T_\parallel = 4$ and $\bar{M}_{biM} \simeq 0.097$ for $T_\perp/T_\parallel = 1/4$. The red line employs a logarithmic horizontal scale using the top axis over a much broader range of $T_\perp / T_\parallel$. The plot shows that $\bar{M}_{biM}$ remains finite even for small or large $T_\perp / T_\parallel$.

It is important to emphasize that the result for the non-Maxwellianity is dependent on the velocity space grid, so Eqs.~(\ref{eq:mbar_bim_grid1}) and (\ref{eq:nonmax_bim_limits}) are only valid when the velocity space grid is proportional to the thermal speed. However, it is typically not practical to have a velocity space grid that varies with temperature. Satellite instrumentation, simulation grids, and laboratory diagnostics typically have a velocity space resolution that is set by other constraints and does not vary in position or time.  For such cases, the general expression in Eq.~(\ref{eq:mbar_bim}) must be used for a bi-Maxwellian distribution function, or Eq.~(\ref{eq:newmbar2}) for an arbitrary distribution function, and care must be taken to ensure $\Delta^3v$ is chosen to properly resolve velocity space structures and preserve a good statistical description.

\section{Numerical Simulations}
\label{sec:sim_setup}

\subsection{The Code and Simulation Setup}
\label{sec:sim_setup_code}

In the following section, we use collisionless particle-in-cell simulations of magnetic reconnection to calculate the non-Maxwellianity measures discussed in the previous sections. Here, the numerical simulation setup is discussed.  The simulation employed here is the same simulation used in \citet{Liang19}, referred to there as the ``base'' simulation.  Only the most relevant details are provided here and the reader is referred to that study for further details. 

The code in use is P3D \citep{zeiler:2002}. The simulations are two-dimensional (2D) in position space and 3D in velocity space. Spatial boundary conditions are periodic in both directions.  Distances are normalized to  the ion inertial scale $d_{i0}$ based on a reference density $n_0$ that is the peak density of the initial current sheet population, magnetic fields are normalized to the upstream field strength $B_0$, velocities are normalized to the Alfv\`en speed $v_{A0}$ based on $B_0$ and $n_0$, times are normalized to the inverse ion cyclotron frequency $\Omega_{ci0}^{-1}$ based on $B_0$, temperatures are normalized to $m_i v_{A0}^2 / k_B$, entropy densities are normalized to $k_B n_0$, and velocity distribution functions are normalized to $n_{0} v_{A0}^3$.

The simulation domain is $51.2 \times 25.6$, and there is a double current sheet configuration with initial half-thickness of 0.5. A uniform background population has density $0.2$, so the ion inertial scale $d_i$ based on the background population is 2.24 $d_{i0}$.  The initial electron and ion populations are drifting Maxwellians, with temperatures of 1/12 and 5/12, respectively, the speed of light is 15, and the ion-to-electron mass ratio is 25.  The grid scale is 0.0125, the time step is 0.001, and the initial number of weighted particles per grid cell is 100; each was chosen to reduce numerical error.  The velocity space grid for kinetic entropy and distribution function calculations is $1 \ v_{A0} \approx 0.69 \ v_{th0,e}$, where $v_{th0,e}$ is the initial electron thermal speed; this choice was justified in \citet{Liang19} and \citet{Liang20}, where results for the total entropy and local entropy density were measured as a function of the velocity space grid and the results were best when the velocity space grid was slightly smaller than the species thermal speed.

\subsection{A Subtlety in Numerically Calculating non-Maxwellianities}
\label{sec:sim_setup_subtlety}

There is an important numerical subtlety concerning the comparison of kinetic entropy density $s$ based on the local distribution function $f(\vec{v})$ with the kinetic entropy density $s_M$ based on a Maxwellianized distribution function $f_M(\vec{v})$. This is because the local distribution function consists of a finite number of macro-particles that is typically relatively small, so there is noise associated with the Monte Carlo approach of the PIC technique.  We find that if one numerically calculates for $n,\vec{u}$, and $T$ for the local distribution $f(\vec{v})$ and constructs an analytical Maxwellian $f_M(\vec{v})$ using these values, the deviation from the local kinetic entropy density is enhanced because $f(\vec{v})$ has PIC noise because it is represented by few particles and $f_M(\vec{v})$ is effectively represented by an infinite number of particles. 

To avoid this undesirable disconnect, one could simply perform the simulation with a larger number of macroparticles per grid. Alternately, one can generate the Maxwellianized kinetic entropy density $s_M$ using a separate Monte Carlo Maxwellian distribution function $f_M(\vec{v})$ using the same number of macro-particles as the local distribution function $f(\vec{v})$; we find this makes the comparison more accurate.  

To do so, we create a Maxwellian entropy density look-up table. The minimum and maximum number of macro-particles $N$ in the position space grid cells of interest is found.  $N$ is proportional to the density $n$ when the macro-particles are unweighted or the variation of the particle weights is very small in the given area. For a number of values in the range of $N$, Maxwellian distributions with a range of $n$ and $T$ are generated, and $s_M$ is calculated for each.  Then, when $n$ and $T$ are found at a particular position space grid cell of interest, the entropy of the Maxwellianized distribution is found by interpolating to that $n$ and $T$ in the look-up table for the corresponding $N$.  The limited number of macro-particles leads to fluctuations of $s_M$ for even the same or similar $n$ and $T$. To reduce the fluctuations, we repeat the Monte Carlo generation of the Maxwellian distribution for each $n$ and $T$ four times, and then smooth the look-up table by averaging over the four. 

There is a caution for using a look-up table when particles carry differing numerical weights. The reason is that the look-up table assumes each macroparticle carries the same weight.  To address this, one can search all the spatial grids in an area of interest in the simulation box to find the variation of the particle weight in this area. If the variation is very small, then the assumption that the number of macro-particles is proportional to the density is reasonable so that the look-up table as a function of $n$ and $T$ is sufficient.  On the other hand, in regions with large variation of particle weight, errors are introduced due to the look-up table. A three-dimensional look-up table as a function of $n$, $T$, and the number of macro-particles $N$ would be needed; this is not carried out for the present study.

\section{Results}
\label{sec:simResults}

\subsection{Validation of the Numerical Implementation}
\label{sec:validation}

Here, we validate the numerical implementation of the Kaufmann and Paterson non-Maxwellianity measure $\bar{M}_{KP}$ discussed in Section~\ref{sec:non_max_KP}.  The most basic metric for validation is whether the kinetic entropy density $s$ and non-Maxwellianity $\bar{M}_{KP}$ give the expected values when the distribution function is Maxwellian.  In the simulation described in Sec.~\ref{sec:sim_setup}, the distributions far upstream from the reconnecting region are essentially Maxwellian throughout the simulation, so we first check to see the numerically generated values are as expected.  

To do so, the local kinetic entropy density $s$ is calculated at every spatial cell using the techniques discussed in \citet{Liang19}.  The associated entropy density $s_M$ of the Maxwellianized distribution function is calculated using a look-up table as described in Section~\ref{sec:sim_setup_subtlety}.  From these, the non-Maxwellianity $\bar{M}_{KP}$ is calculated using Eq.~(\ref{eq:flnf_M}).

\begin{figure}
\includegraphics[width=3.4in]{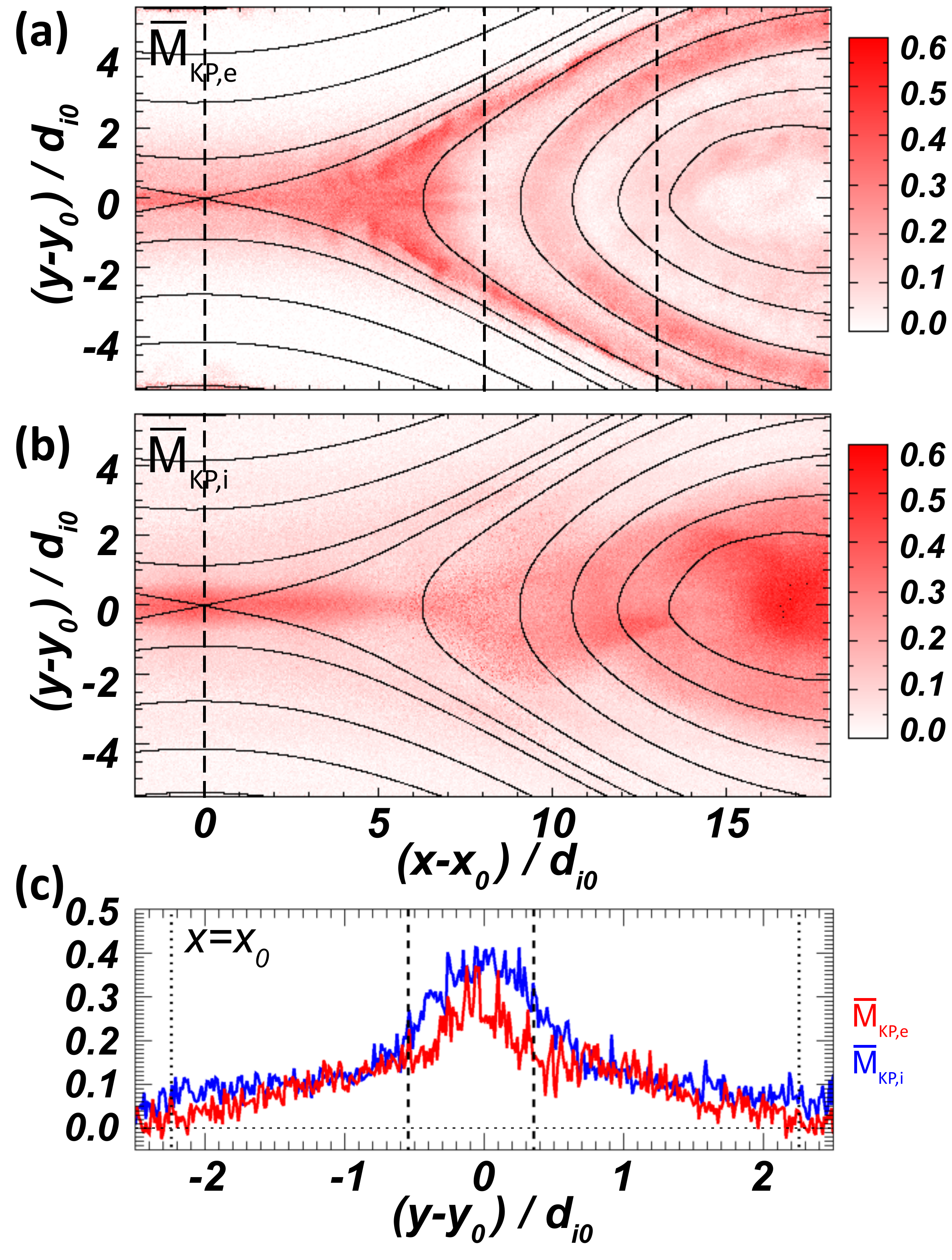}
\centering
\caption{2D plots of the Kaufmann and Paterson non-Maxwellianity at $t = 41$, where the location of the X-point at this time is $(x_0,y_0)$, for (a) electron $\bar{M}_{KP,e}$ and (b) ions $\bar{M}_{KP,i}$.  Black solid lines are magnetic field lines. Vertical dashed lines in (a) at $x-x_0$ = 0, 8, and 13 indicate cuts that are investigated further in Figs.~\ref{fig5:distcuts0} - \ref{fig7:distcuts13}. (c) Vertical cuts of $\bar{M}_{KP,e}$ (red) and $\bar{M}_{KP,i}$ (blue) through the X-point in the $y$ direction.  The vertical dashed and dotted lines mark the edges of the non-frozen-in region for electrons and ions, respectively.} \label{fig4:mbarei}
\end{figure}
The Kaufmann and Paterson non-Maxwellianity $\bar{M}_{KP}$ at time $t = 41$ is given in Fig.~\ref{fig4:mbarei}, zoomed in to a portion of the computational domain near the X-point and in the outflow region for (a) electrons and (b) ions, respectively.  The coordinate system is relative to the location of the X-point $(x_0,y_0)$, so that the X-point is at the origin in these plots. A vertical cut through the X-point is shown in panel (c) for electrons (red) and ions (blue).  For both electrons and ions, the non-Maxwellianity is near zero in the upstream regions (beyond $1 \ d_i \simeq 2.24 \ d_{i0}$ upstream of the reconnection site), as desired.  When we use an entropy density of the Maxwellianized distribution function based on the analytical value for the fluid density, bulk flow, and effective temperature from the simulation instead of the look-up table, the kinetic entropy density $s_M$ within the current sheet is slightly lower than the kinetic entropy density from the look-up table by about 0.01 to 0.04 (about 5\%) for the simulation performed here.

Looking holistically at the rest of the domain, the value of $\bar{M}_{KP}$ is mostly non-negative for both species, as expected from  Section~\ref{sec:basicpropmbarkp}. Numerical effects due to the finite number of macro-particles in the simulation lead to fluctuations and potentially small negative values for $\bar{M}_{KP}$.  This suggests the implementation of the kinetic entropy in the simulations and the look-up table is valid, and underscores the importance of using the look-up table for the entropy of the Maxwellianized distribution function for the number of macroparticles per grid in use in the present simulations. 

\subsection{Interpreting the Non-Maxwellianity}
\label{sec:resultsinterp}

We now revisit Fig.~\ref{fig4:mbarei} to investigate the non-Maxwellianity of the plasma in the regions affected by reconnection, with a goal of relating the non-Maxwellianity to known physical processes in reconnection.  Panels (a) and (b) show that both electrons and ions depart significantly from zero non-Maxwellianity in the diffusion region $-5 < x - x_0 < 5$ and $-2 < y - y_0 < 2$. Both also depart significantly from zero in the reconnection exhaust $5 < x - x_0 < 8$.  There is a magnetic island roughly at $x-x_0>8$.

Panel (a) reveals that $\bar{M}_{KP,e}$ for the electrons has its largest departures from zero in the diffusion region, along the separatrices, and at the boundary of the island.  In contrast, panel (b) shows that $\bar{M}_{KP,i}$ departs from zero in the diffusion region and in the core of the island.  These regions are consistent with where we expect electrons and ions to be non-Maxwellian.  Electrons accelerate due to the reconnection electric field and undergo non-adiabatic motion in the electron diffusion region, form counter-streaming beams and electron holes near separatrices, and are Fermi-accelerated and heated in the magnetic island \citep{Drake06}.  For ions, they undergo acceleration and non-adiabatic motion in the ion diffusion region, form counter-streaming beams and pickup-ion acceleration in the exhaust \citep{Drake09a}, and are reflected by the moving jet front in the magnetic islands. The physical picture will be confirmed for electrons by investigating distribution functions in Section~\ref{sec:resultsdist}.  

The vertical cuts of non-Maxwellianity through the X-point shown in Fig.~\ref{fig4:mbarei}(c) more clearly shows that the departure from zero occurs for both species near $1 \ d_i \simeq 2.24 \ d_{i0}$ upstream from the X-point.  By inspecting traces of the ion inflow velocity $v_{iy}$ and $(\vec{E} \times \vec{B})_y / B^2$ (not shown), we find the ion bulk inflow deviates from the $\vec{E} \times \vec{B}$ drift speed at approximately $|y - y_0| \simeq 2.25 \simeq 1 \ d_i$. This deviation defines the upstream edge of the ion diffusion region and is denoted by the two vertical dotted lines in panel (c).  Therefore, the ion non-Maxwellianity $\bar{M}_{KP,i}$ begins to depart from zero at the edge of the ion diffusion region, which is where ions are expected to become demagnetized and therefore their distributions become non-Maxwellian.

\begin{figure}
    \includegraphics[width=3.4in]{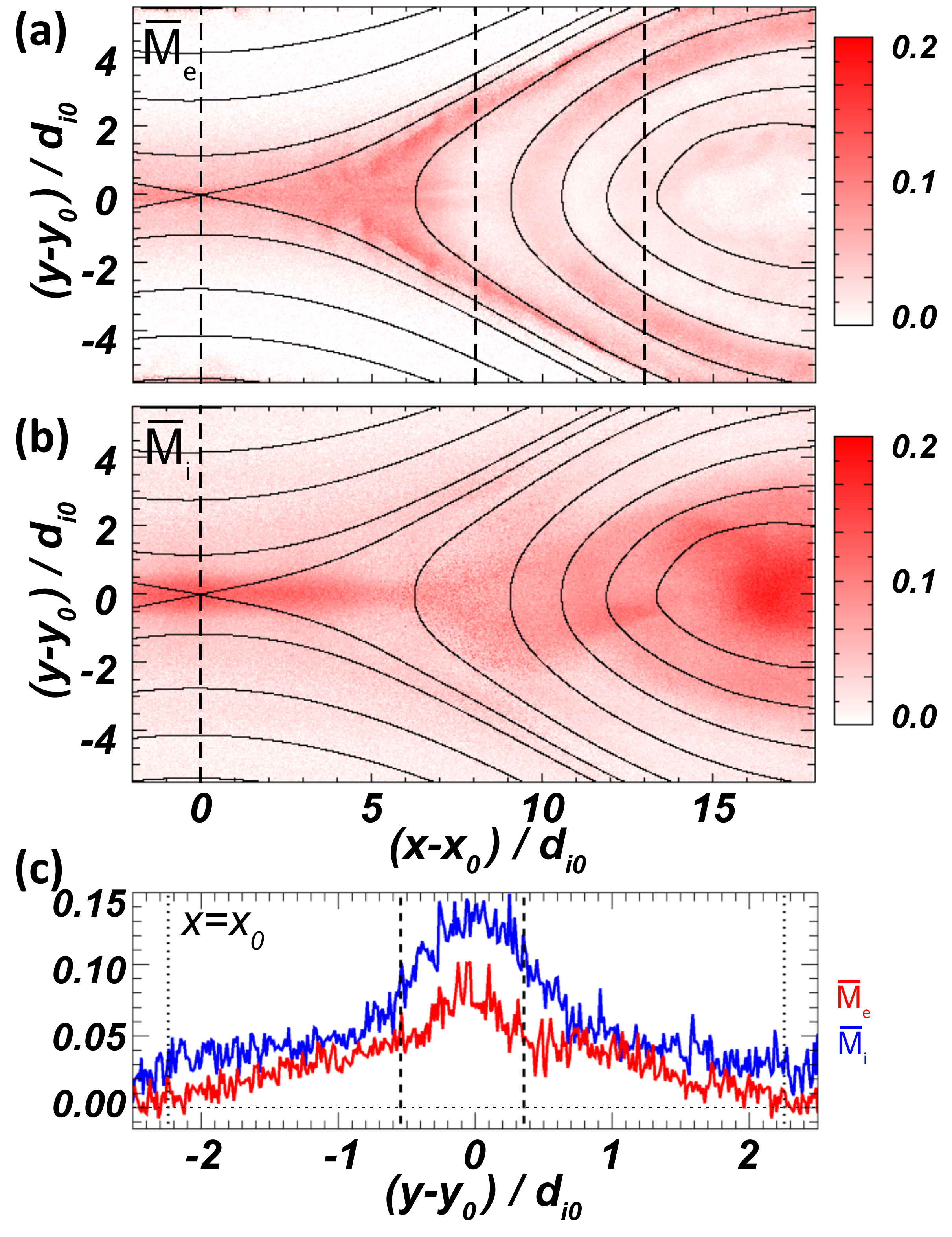}
\centering
\caption{Same as Fig.~\ref{fig4:mbarei} but for the non-Maxwellianity measure $\bar{M}$.} \label{fig4:mbarei2}
\end{figure}
Interestingly, $\bar{M}_{KP,e}$ also departs from 0 starting at the edge of the ion diffusion region, {\it i.e.,} outside the electron diffusion region defined by where $v_{ey}$ differs from $(\vec{E} \times \vec{B})_y / B^2$. As we will show in Section~\ref{sec:resultsdist}, this corresponds to the region of electron trapping upstream of the electron diffusion region discussed extensively by Egedal and colleagues \citep{egedal13}.  Both ion and electron non-Maxwellianity measures increase in magnitude as one approaches the X-point.

The upstream edges of the electron diffusion region are denoted by the vertical dashed lines at $y-y_0 \simeq 0.35$ and $-0.55$ in panel (c); the electron diffusion region therefore has a half-width of $0.45 \ d_{i0} = 2.25 \ d_{e0} \simeq 1 \ d_e $, where $d_{e0}$ and $d_e$ are the electron inertial lengths based on the density $n_0$ and the background plasma density $0.2$, respectively.  Both electrons and ions see larger increases to their non-Maxwellianity in the electron diffusion region.  This suggests that the non-Maxwellianity could be potentially useful as one approach among many to identify reconnection diffusion regions.  

\subsection{The New Non-Maxwellianity Measure}
\label{sec:resultsnewmbar}

Here, we plot the new non-Maxwellianity measure discussed in Sec.~\ref{sec:defmbar}. Data analogous to Fig.~\ref{fig4:mbarei} for $\bar{M}_{KP}$ are plotted in Fig.~\ref{fig4:mbarei2} for $\bar{M}$.  The spatial structure of the two measures are quite similar.  This is to be expected from Eq.~(\ref{eq:newmbar2}), since $\bar{M}$ is proportional to $\bar{M}_{KP}$ and the argument in the denominator is inside a natural logarithm so there is only a weak dependence on temperature.  The range of values for $\bar{M}$ is from 0 to about 0.10 for electrons and 0 to 0.15 for ions.

\subsection{Analysis of Electron Distributions and Non-Maxwellianity}
\label{sec:resultsdist}

Here, we investigate the non-Maxwellianity parameters in relation to distribution functions measured in the simulations.  This allows us to ensure the non-Maxwellianity measures are correctly identifying distributions that are non-Maxwellian, to associate physical features with the non-Maxwellianity, and to quantify non-Maxwellianity measures in the context of the exact solutions in Sections~\ref{sec:ent_nonmaxsub1_beams} - \ref{sec:ent_nonmaxsub1_egedal} and \ref{sec:defmbar}.

\begin{figure}
\centering
\includegraphics[width=3.4in]{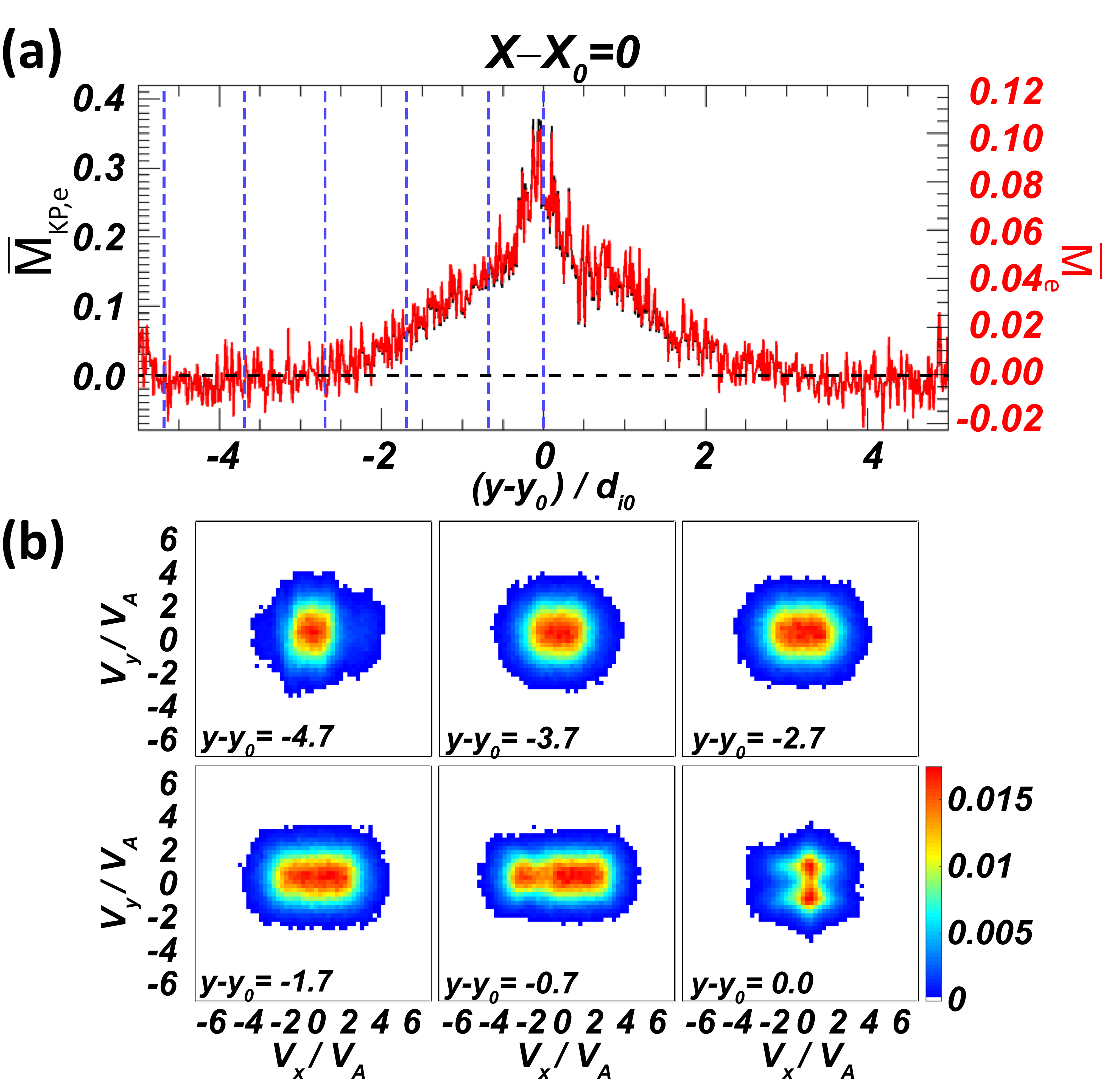}
\caption{(a) Non-Maxwellianity measures $\bar{M}_{KP,e}$ (black) and $\bar{M}_e$ (red) in a vertical slice through $x-x_0=0$ at $t = 41$. (b) Reduced electron velocity distribution functions in the $(v_x,v_y)$ plane at $x-x_0=0$ and $y-y_0=-4.7,-3.7,-2.7,-1.7,-0.7,$ and 0.0, denoted by the vertical dashed lines in panel (a).} \label{fig5:distcuts0}
\end{figure}
\begin{figure}
  \centering
  \includegraphics[width=3.4in]{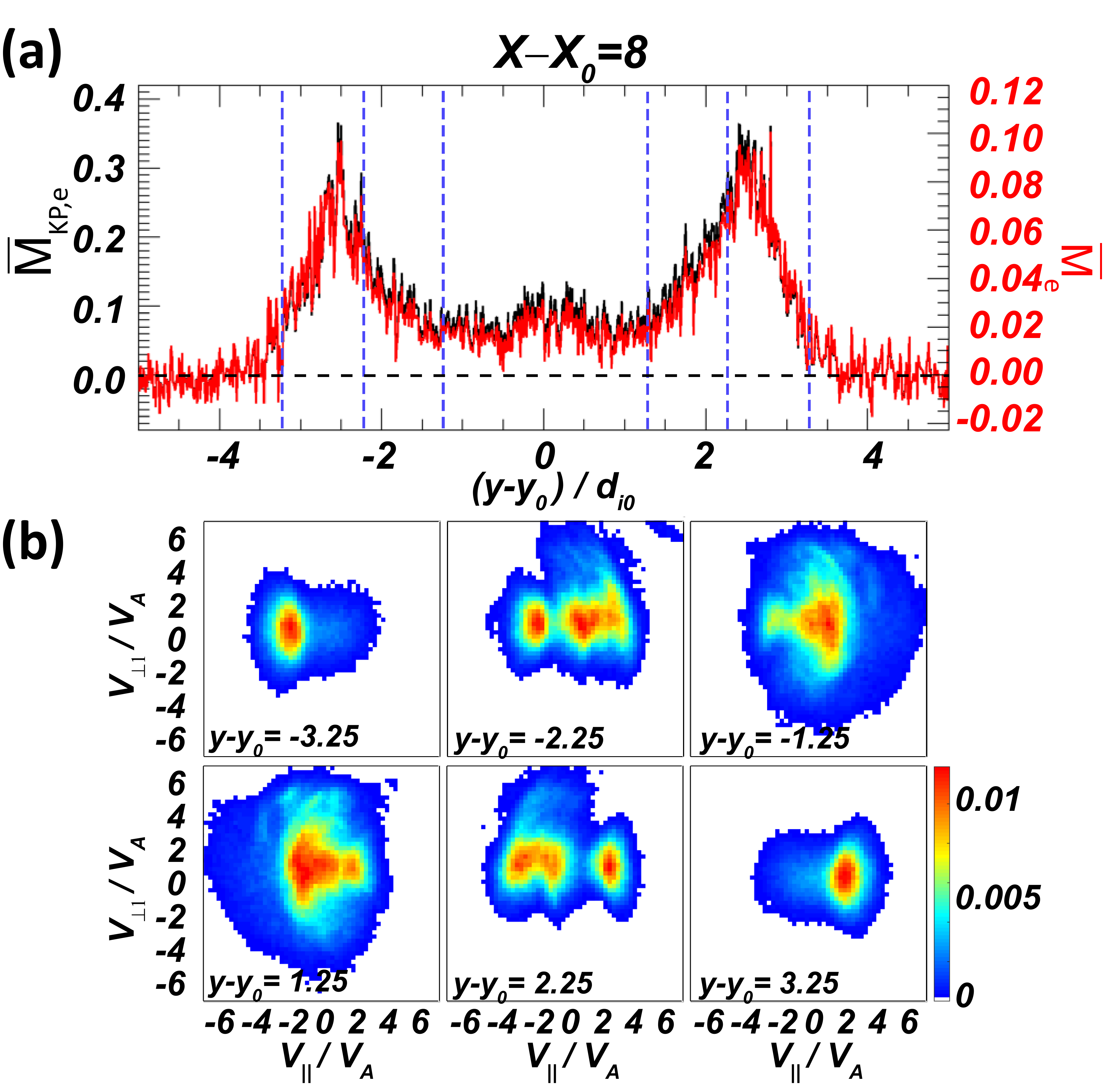}
  \caption{Similar to Fig.~\ref{fig5:distcuts0} except at $x-x_0=8$. The vertical dashed lines in (a), where the velocity distribution functions are plotted in (b), are at $y-y_0=\pm3.25,\pm2.25$, and $\pm1.25$.  Here, the reduced distribution functions are plotted in the ($v_\parallel,v_{\perp1})$ plane.}
\label{fig6:distcuts8}
\end{figure}
\begin{figure}
\centering 
\includegraphics[width=3.4in]{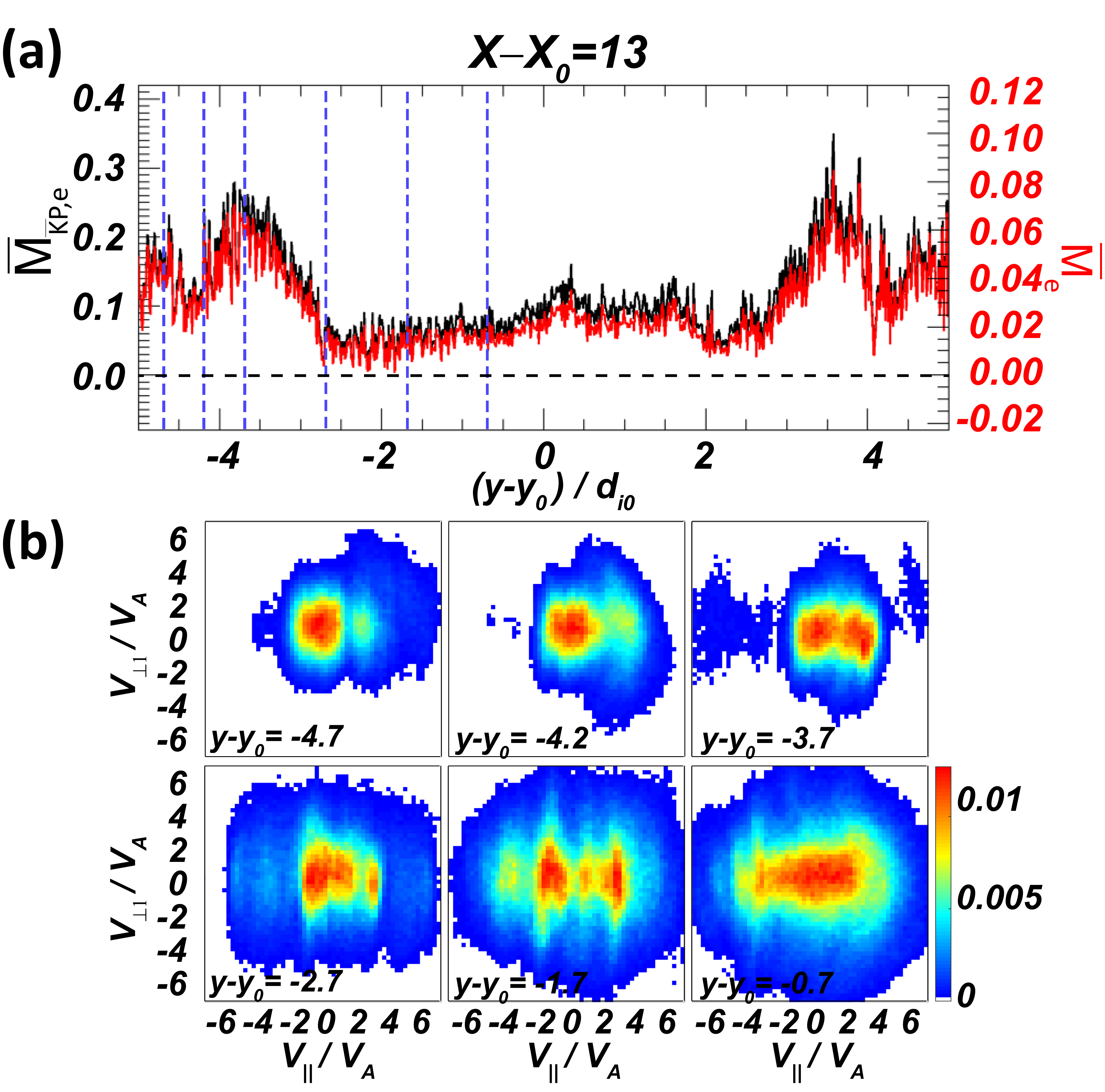}
\caption{Similar to Fig.~\ref{fig6:distcuts8} except at $x-x_0 = 13$.  The locations of the distributions are at $y-y_0=-4.7, -4.2, -3.7, -3.2, -2.7, -1.7,$ and $-0.7$.} \label{fig7:distcuts13}
\end{figure}

We treat distributions at the three cuts denoted by the vertical dashed lines in Figs.~\ref{fig4:mbarei}(a) and~\ref{fig4:mbarei2}(a), which are at $x-x_0=0,8$ and 13.  For brevity, we only treat electrons. Results are shown in Figs.~\ref{fig5:distcuts0} -~\ref{fig7:distcuts13}, respectively. In each figure, panel (a) shows the Kaufmann and Paterson electron non-Maxwellianity $\bar{M}_{KP,e}$ in black and the electron non-Maxwellianity $\bar{M}_e$ in red as a function of $y-y_0$.  Each has six $y-y_0$ positions marked by blue vertical dashed lines.  In panel (b) of each figure, the reduced velocity distribution functions (VDFs), {\it i.e.,} the VDF with the third dimension integrated out, at the six marked positions are plotted.  For $x - x_0 = 0$ (Fig.~\ref{fig5:distcuts0}), the VDFs are plotted in the $(v_{ex},v_{ey})$ plane.  For $x - x_0 = 8$ and $13$ (Figs.~\ref{fig6:distcuts8} and \ref{fig7:distcuts13}), the VDFs are plotted in the $(v_\parallel,v_{\perp 1})$ plane, where $v_\parallel$ is in the local direction of the magnetic field and $v_{\perp 1}$ is in the local $\vec{E} \times \vec{B}$ direction.

We begin with the plots at  $x - x_0 = 0$ in Fig.~\ref{fig5:distcuts0}.  At this $x$ location, except at $y-y_0=0$, the $x$ direction is approximately parallel to the local magnetic field and $y$ is perpendicular to it in the simulation plane.  At $y-y_0=-4.7, -3.7, -2.7$, the non-Maxwellianities $\bar{M}_{KP,e}$ and $\bar{M}_e$ are close to zero and the VDFs resemble isotropic Maxwellian distribution functions, as is expected to be the case outside the ion diffusion region.  

\begin{table}
  \begin{center}
\def~{\hphantom{0}}
  \begin{tabular}{llccccc}
      Name & $(x-x_0,y-y_0)$ & Input from simulations & $\bar{M}_{KP,e}^{pred} $ & $\bar{M}_e^{pred}$ & $\bar{M}_{KP,e}^{sim}$ & $\bar{M}_e^{sim}$ \\[3pt]
      Egedal & (0,-0.7) & $n = 0.15, n_\infty = 0.186, \phi_\parallel = 0.070$, & 0.151~~ & 0.0430 & 0.13~ & 0.037 \\
      & & $T_\infty = 0.08, B_\infty = 0.51, B = 0.29,$ & & & & \\
       & & $\Rightarrow n/n_\infty = 0.805, b = 0.57, \Phi = 0.88$ & & & & \\
      Beam & $(8,2.25)$ & $n_1 = 0.0457, u_{z1} = 2.455, T_1 = 0.0350,$ & 0.323~ & 0.0868 & 0.26~ & 0.068 \\
       & & $n_2 = 0.114, u_{z2} = -1.80, T_2 = 0.0534$& & & & \\
       Bi-M & $(13,-0.7)$ & $T_\perp = 0.197, T_\parallel = 0.425$  & 0.0700 & 0.0147 & 0.079 & 0.017 \\
  \end{tabular}
  \caption{Comparison of analytic non-Maxwellianities to simulation results. The first column gives the name of the distribution, and the second gives the location of the simulated distribution (Figs. \ref{fig5:distcuts0}-\ref{fig7:distcuts13} for $x - x_0 = 0, 8,$ and 13, respectively. The third column gives the necessary plasma parameters for the given analytic distribution extracted from the simulations, given in normalized code units. The fourth and fifth give the predicted $\bar{M}_{KP,e}^{pred}$ and $\bar{M}_e^{pred}$ for each model, and the sixth and seventh give the measured values $\bar{M}_{KP,e}^{sim}$ and $\bar{M}_e^{sim}$ from the simulations.}
  \label{tab:mbar}
  \end{center}
\end{table}
At $y-y_0=-1.7$ and $-0.7$, inside the ion diffusion region, $\bar{M}_{KP,e}$ and $\bar{M}_e$ are non-zero.  At these locations, the distributions are elongated in the direction parallel to the magnetic field. This is due to trapped electrons just upstream of the electron diffusion region \citep{egedal13}, as discussed in Section~\ref{sec:ent_nonmaxsub1_egedal}. A quantitative comparison with predictions from that section would be desirable, but it requires having asymptotic values in a region with Maxwellian distributions.  This is not achieved in the small simulation carried out here.  However, we estimate the value as best we can for this simulation.  For the $y-y_0 = -0.7$ case, we start by integrating the parallel electric field along the magnetic field from $x_0$ to the point where $E_\parallel$ becomes negative.  We have to stop the integration at this point because the simulation domain is not large enough to reach a plasma that is Maxwellian as one goes out along the magnetic field line.  Proceeding anyway, we estimate the relevant necessary input parameters from the simulation at $(x-x_0,y-y_0) = (0,-0.7)$, which are collected in the first row in Table~\ref{tab:mbar}.  Using Eqs.~(\ref{I4def}) - (\ref{L567def}), we get $s_{Eg} \approx 1.04$, and using Eqs.~(\ref{Tparalocaldef}) - (\ref{TlocalEg}), we get $T_{Eg} = 0.078$. Then, the predicted non-Maxwellianities $\bar{M}_{KP,e}^{pred}$ and $\bar{M}_e^{pred}$ are calculated from Eqs.~(\ref{eq:mbaregedal}) and (\ref{eq:newmbar2}), and are provided in Table~\ref{tab:mbar}. Despite the simulation size being too small, the prediction agrees within 20\% of the simulated values $\bar{M}_{KP,e}^{sim}$ and $\bar{M}_e^{sim}$ at $y - y_0 = -0.7$ in the simulation as seen in Fig.~\ref{fig5:distcuts0}(a) and given numerically in the table. As one goes further away in $y$ from the X-point, the agreement gets worse, which is a result of our system size not being large enough.

At $y-y_0=0$, the electrons form a beam in the out-of-plane $z$ direction (not shown) due to being accelerated by the reconnection electric field as they undergo meandering motion [{\it e.g.,} \citep{Ng11}]. The VDF is significantly different than an isotropic Maxwellian.  Indeed, this point is the local maximum of the non-Maxwellianity $\bar{M}_{KP,e}$ and $\bar{M}_e$ in panel (a). 

We turn to $x-x_0 = 8$, shown in Fig.~\ref{fig6:distcuts8}. At this $x$ location, the separatrices are at $y-y_0 \simeq \pm3$, so $|y-y_0|<3$ is the exhaust and $|y-y_0|>3$ is in the upstream region. The VDFs at $y-y_0=\pm3.25$ show inflowing cold electron beams anti-parallel ($y-y_0 = -3.25$) and parallel ($y-y_0 = 3.25$) to the magnetic field as they convect in towards the X-point. The VDFs at $y-y_0= \pm1.25$ show the hotter electron beams in the exhaust flowing in the opposite direction inside the separatrix. These flows are the well-known Hall currents that result in the Hall magnetic field \citep{Sonnerup79}. The non-Maxwellianity in these locations is non-zero, but is relatively low because the VDF does not differ much from a Maxwellian. 

At $y-y_0=\pm2.25$, between these two locations, there are counter-streaming signatures due to populations from both beams. While the two beams are not totally separated in velocity space, we check to see if the non-Maxwellianity is reasonably well reproduced by the analytical prediction for two beams from Section~\ref{sec:ent_nonmaxsub1_beams}. For the distribution at $(x-x_0,y-y_0) = (8,2.25)$, we take a cut along $v_{\perp 1} = 0$ and fit the two populations with Maxwellians. The resulting plasma parameters are given in the second row of Table~\ref{tab:mbar}. From Eq.~(\ref{eq:tmbeam}), $T_{beam} = 0.0974$, and Eq.~(\ref{eq:mbarkpbeam}) gives  $\bar{M}_{KP,e}^{pred}$ given in the table, along with the associated $\bar{M}_{e}^{pred}$ from Eq.~(\ref{eq:newmbar2}). The measured values $\bar{M}_{KP,e}^{sim}$ and $\bar{M}_e^{sim}$ are given in the table; they are (within 22\%) of the predicted values for non-overlapping counter-streaming beams, which is reasonably close.

Finally, we look further downstream cutting mostly through the interior of a magnetic island at $x-x_0 =13$ in Fig.~\ref{fig7:distcuts13}. At $y-y_0=-4.7$ and $-4.2$, the VDFs display signatures of counter-propagating beams along the magnetic field, similar to $y-y_0=-2.25$ in Fig.~\ref{fig6:distcuts8}.  The non-Maxwellianity is elevated there.

At $y-y_0=-3.7$, the VDF is in a region near the separatrix of a secondary X-point in the outflow that had formed before $t=41$.  Multiple electron beams are visible, due to the bouncing of electrons in the island. They lead to a relatively large value of $\bar{M}_{KP,e}$ and $\bar{M}_e$. Closer to the neutral line, as $y-y_0$ goes from $-2.7$ to $-1.7$ to $-0.7$, the multiple beams have thermalized and form hot electron distributions that are elongated along the parallel direction. This is a signature of Fermi acceleration in a contracting island \citep{Drake06}. Assuming the VDF is similar in both perpendicular directions, we check to see if the analytical prediction of $\bar{M}_{KP}$ for a bi-Maxwellian distribution is in reasonable agreement with the simulation result.  For the distribution at $(x-x_0,y-y_0) = (13,-0.7)$, we fit Maxwellians to the $v_\parallel = 0$ and $v_{\perp 1} = 0$ cuts of this VDF; the resulting $T_\perp$ and $T_\parallel$ are in the third row of Table~\ref{tab:mbar}.  Using Eqs.~(\ref{eq:mbarkpbimax}) and (\ref{eq:mbar_bim}), the predicted non-Maxwellianities $\bar{M}_{KP,e}^{pred}$ and $\bar{M}_{e}^{pred}$ are shown in the table. The simulated values $\bar{M}_{KP,e}^{sim}$ and $\bar{M}_e^{sim}$ are shown in the table, as well, showing agreement with the predicted values within 10-20\%.

In summary, the non-Maxwellianity measures $\bar{M}_{KP,e}$ and $\bar{M}_e$ can be reliably implemented as diagnostics in kinetic PIC simulations. These local measures can be plotted as function of space and time, and are capable of easily identifying where distribution functions are non-Maxwellian, both for electrons and ions.  For magnetic reconnection, the locations of elevated non-Maxwellianity coincide with kinetic-scale physical processes related to particle acceleration and non-adiabatic particle motion. The velocity distribution functions in regions of non-zero simulated non-Maxwellianity are, indeed, non-Maxwellian, and the analytic calculations in Section~\ref{sec:ent_nonmax_kp_theory} are capable of motivating the relative size of the non-Maxwellianity in some regions within about 20\% despite a variation of the non-Maxwellianities by over a factor of 4.

\section{Discussion and Conclusions}
\label{sec:disc_conc}

We investigate a number of aspects of kinetic entropy-based measures of the non-Maxwellianity of a given distribution function $f(\vec{v})$. The first, which we call $\bar{M}_{KP}$ and is given in Eq.~(\ref{eq:flnf_M}), was developed by \citet{Kaufmann09} and used to analyze observational data of Earth's plasma sheet. Their measure is the difference between the local kinetic entropy density $s = - k_B \int d^3v f(\vec{v}) \ln f(\vec{v})$ and the kinetic entropy density $s_M$ of the Maxwellianized distribution function $f_M(\vec{v})$ based on the low order fluid moments of the full distribution function $f(\vec{v})$, and normalized to the number density and the specific heat per unit volume for an ideal gas.  As stated by Kaufmann and Paterson, this non-Maxwellianity is a good measure because it is non-negative and only vanishes when $f(\vec{v})$ is Maxwellian.  Moreover, when collisions are present, regions of non-Maxwellianity are regions in which collisions are expected to be important, so the locations of elevated non-Maxwellianity are also those in which irreversible dissipation is more prone to occur [{\it e.g.,} \citep{Pezzi16,Pezzi19}].

We are unaware of previous work to develop a quantitative understanding of $\bar{M}_{KP}$, so in this study we derive closed-form analytical expressions for the non-Maxwellianity for common non-Maxwellian distributions, including two parallel or anti-parallel beams well-separated in velocity space (Section~\ref{sec:ent_nonmaxsub1_beams}), bi-Maxwellian distribution functions for anisotropic plasmas (Section~\ref{sec:ent_nonmaxsub1_bimax}), and Egedal distributions that arise near magnetic reconnection sites (Section~\ref{sec:ent_nonmaxsub1_egedal}).

In addition, we show that there are undesirable features of $\bar{M}_{KP}$. The measure can diverge in various physical limits (especially related to temperature going to zero or infinity). This makes it difficult to interpret what a particular numerical value for $\bar{M}_{KP}$ means. We argue that the reason for this is that the measure is based on the kinetic entropy density, which beyond being the kinetic entropy per unit volume does not have a physical interpretation of entropy, and it does not even have the units of entropy density. Rather, the velocity space entropy density \citep{Mouhot11,Liang19} has appropriate units and a physical interpretation of the number of ways to exchange the velocities of particles in the distribution in velocity space at a given position and time, thus being a more physically meaningful measure of the local kinetic entropy density associated with a given distribution function $f(\vec{v})$.

We introduce a new non-Maxwellianity measure $\bar{M}$ based on the velocity space kinetic entropy density. We show it is proportional to $\bar{M}_{KP}$ [see Eq.~(\ref{eq:newmbar2})], with a denominator that regularizes the measure because it has explicit dependence on the velocity space grid scale.  This dependence is not a hindrance, but rather is a feature, as it captures the physical effect that the kinetic entropy does depend on the velocity space grid scale.

We then use collisionless particle-in-cell simulations of two-dimensional anti-parallel magnetic reconnection, the same simulation studied in detail in \citet{Liang19}, to study the non-Maxwellianity measures numerically. We validate the numerical implementation of the non-Maxwellianity measure. When the number of particles per grid cell is not exceedingly high, we find that it is important to use a look-up table for the entropy density of the Maxwellianized distribution function so that it has the same level of numerical error as the raw distributions.  We show that, for the simulation considered here, the non-Maxwellianity measure is non-zero where kinetic-scale processes drive velocity distribution functions away from Maxwellianity.  We also show the analytic calculations in Section~\ref{sec:ent_nonmax_kp_theory} give reasonable agreement with appropriate naturally occurring distributions, within about 20\% of the numerically calculated non-Maxwellianity.  

We argue that the non-Maxwellianity, in concert with other measures, can be useful to help interpret satellite observations, laboratory experiments, and kinetic numerical simulations. We suggest that $\bar{M}_{KP,e}$ and $\bar{M}_e$ could be useful to identify and evaluate the dynamics of plasma dissipation, such as regions of interest like electron and ion diffusion regions in reconnection, intermittent current sheets in turbulence, and boundary layer physics in collisionless shocks (provided the resolution of the measurements is high enough to resolve kinetic-scale structures). These local measures can be plotted as function of space and time, and are capable of identifying spatial regions where distribution functions are non-Maxwellian, both for electrons and ions.  Now that satellites routinely measure plasma distribution functions at very high resolution, measuring non-Maxwellianity in real data is achievable, as was shown in previous studies of the enstrophy non-Maxwellianity using Magnetospheric Multiscale (MMS) mission data \citep{Servidio17}.  This will be undertaken for entropy-based non-Maxwellianity measures in future work.  To get a feel for the required sensitivity, the simulation here suggests that $\bar{M}_{KP,e}$ is up to about 0.6, corresponding to a difference of kinetic entropy per particle $(s_M-s)/n$ of $1.2 \times 10^{-23}$ J/K $= 7.8 \times 10^{-5}$ eV/K.

It is worth discussing the results of the present study in light of the observational results in \citet{Kaufmann09}.  In their observational study, Geotail data from ten years of ion measurements in one minute bins was used and averaged, and binned over spatial regions of Earth's plasma sheet. Their results reveal non-Maxwellianities (their Fig.~5) on the order of 1.4-2 throughout the entire plasma sheet. These values greatly exceed the values for $\bar{M}_{KP,i}$ observed in our reconnection simulations [our Fig.~\ref{fig4:mbarei}].  There are numerous reasons for the significant difference. One is that the parameters of our simulation and the parameters in the ten year study may not be commensurate. Another potential issue is that their fluid entropy was calculated from the fluid variables using the analytic expression, and this could cause an offset in the manner discussed in Sec.~\ref{sec:sim_setup_subtlety}, which would inflate the measured non-Maxwellianity.  In addition, the Geotail data came from one minute averages, which undoubtedly captured variations in plasma parameters such as density and temperature, which would introduce non-Maxwellianities that may not be present from a higher time resolution measurement. Satellite data also often does not capture the cold plasma population, which requires assumptions for calculating entropy and may introduce uncertainties.  Using, for example, MMS data, will be a more direct comparison with the simulation results presented here, and will be pursued in future work.

While the non-Maxwellianity is a potentially useful measure to aid the interpretation of data and simulations, a number of aspects should be kept in mind. First, a non-zero non-Maxwellianity can indicate reversible and irreversible processes, {\it i.e.,} it is not necessarily purely irreversible.  Moreover, it is true that non-Maxwellianity identifies regions where irreversible dissipation is prone to occur through collisions, but if collisions are entirely absent then the non-Maxwellianity is not associated with irreversible dissipation. To address this more fully requires comparisons with collisional simulations, which will be addressed in the companion study \citep{Pezzi20}.  Also, since the non-Maxwellianity is local in position space, it is not capable of identifying if a plasma element has undergone an entropy change as it evolves, {\it i.e.,} in a Lagrangian reference frame.

An ambiguity of the non-Maxwellianity is that there is not a one-to-one correspondence between a particular value of it and an associated distribution function or physical process.  Rather, to understand it quantitatively, one still needs information about the structure of the distribution function, such as whether it consists of beams or bi-Maxwellian plasmas.  However, once that link is established, the value of $\bar{M}$ can give perspective about the plasma, including allowing the inference of quantities such as the temperature anisotropy. In addition, it provides information about what fraction of the kinetic entropy density a given distribution function has given up relative to its associated Maxwellian value, which has the maximum kinetic entropy for a fixed number of particles and total energy.  Further work on interpreting such results would be worthwhile.  On the other hand, there are independent efforts underway to identify structures in complicated distribution functions by breaking them into separate populations [{\it e.g.}, \citep{Goldman20}], which could be used to aid in the interpretation of the entropy-based results.

The kinetic entropy density and non-Maxwellianity should be used in concert with other diagnostics and measures of plasma processes to contribute to its interpretation. A number of other measures have been developed; the non-Maxwellianity and these other measures will be compared and contrasted in simulations of magnetic reconnection and plasma turbulence in collisionless and collisional kinetic simulations in a companion study \citep{Pezzi20}.

\acknowledgments

This manuscript was prepared for a special issue associated with Vlasovia 2019: The Sixth International Workshop on the Theory and Applications of the Vlasov Equation.  We acknowledge many helpful conversations at the workshop.  We also acknowledge some support on numerical aspects with Benjamin Woods and Raju Bhai KC at WVU, and helpful discussions and a numerical inversion of Eq.~(\ref{I4def}) from J.~Egedal. The authors thank Francesco Pecora, Denise Perrone, Vadim Roytershteyn, and Luca Sorriso-Valvo for helpful comments on the manuscript. Data for the plots in this paper are available at doi:10.5281/zenodo.3951730. HL and GPZ acknowledge support from NSF EPSCoR RII-Track-1 Cooperative Agreement OIA-1655280 and an NSF/DOE Partnership in Basic Plasma Science and Engineering via NSF grant PHY-1707247. HL, MHB, and PAC gratefully acknowledge support from NSF Grant PHY-1804428, NASA Grants NNX16AG76G and 80NSSC19M0146, and DOE Grant DE-SC0020294. SS and FV acknowledge funding from the European Union Horizon2020 research and innovation programme under grant agreement No. 776262 (AIDA, www.aida-space.eu).  This research uses resources of the National Energy Research Scientific Computing Center (NERSC), a DOE Office of Science User Facility supported by the Office of Science of the U.S.~Department of Energy under Contract No.~DE-AC02-05CH11231.

%

\end{document}